\documentclass[useAMS,usenatbib,usegraphicx]{mn2e}

\begin{document}
\newcommand{\Msun}{M_\odot}
\newcommand{\be}{\begin{equation}}
\newcommand{\ee}{\end{equation}}
\newcommand{\sdssu}{u^*}
\newcommand{\sdssg}{g^*}
\newcommand{\sdssr}{r^*}
\newcommand{\sdssi}{i^*}
\newcommand{\sdssz}{z^*}
\newcommand{\sdssri}{r^* - i^*}
\newcommand{\sdssug}{u^* - g^*}
\newcommand{\sdssgr}{g^* - r^*}
\newcommand{\Mr}{M_{r^*}}
\newcommand{\Vmax}{$V_{max}$}
\newcommand{\masyr}{mas yr$^{-1}$}
\newcommand{\kms}{km s$^{-1}$}
\newcommand{\Ic}{I_{\rm C}}

\title[The Subdwarf Luminosity Function]{The subdwarf luminosity function}

\author[A.P. Digby et al..]  {A.P. Digby$^{1}$\thanks{Email: apd@roe.ac.uk}, N.C. Hambly$^{1}$,
J.A. Cooke$^{1}$, I.N. Reid$^{2}$, R.D. Cannon$^{3}$ \\
$^{1}$Institute for Astronomy, University of Edinburgh, Royal
Observatory, Blackford Hill, Edinburgh, EH9 3HJ, UK\\ $^{2}$Space
Telescope Science Institute, 3700 San Martin Drive, Baltimore, MD
21218, USA\\ $^{3}$Anglo-Australian Observatory, PO Box 296, Epping, NSW
2121, Australia}

\date{Accepted 2003 May 24. Received 2003 May 20; in original form
2003 March 31}

\maketitle

\begin{abstract}

Using data from the Sloan Digital Sky Survey Early Data Release and
SuperCOSMOS Sky Survey scans of POSS-I plates, we identify a sample of
$\sim$2600 subdwarfs using reduced proper motion methods and strict
selection criteria. This forms one of the largest and most reliable
samples of candidate subdwarfs known, and enables us to determine
accurate luminosity functions along many different lines of sight. We
derive the subdwarf luminosity function with unprecedented accuracy to
$M_V \la$ 12.5, finding good agreement with recent local estimates but
discrepancy with results for the more distant spheroid. This provides
further evidence that the inner and outer parts of the stellar halo
cannot be described by a single density distribution. We also find
that the form of the inner spheroid density profile within
heliocentric distances of 2.5 kpc is closely matched by a power law
with an index of $\alpha = -3.15 \pm 0.3$.
\end{abstract}

\newpage


\section{Introduction} \label{sec:intro}

As the most common members of the spheroid, subdwarf stars are crucial
to understanding the structure and evolution of Galactic halo. Their
local scarcity means that they are difficult to detect, although
proper motion studies \citep*[eg.][]{Carney et al.1996} have provided
several examples of the brighter, early-type (F-K) subdwarfs in the
solar neighbourhood, yielding important information about their
chemical composition, density and kinematics. However, only small
samples exist of both the fainter, later-type subdwarfs beyond the
immediate solar neighbourhood. As a result, there are significant
uncertainties surrounding the characteristics of the stellar halo at
larger distances and the shape and normalisation of the subdwarf
luminosity function.

The advent of large scale, deep surveys with accurate photometry such
as the SDSS has brought many more of these stars within the
observational reach of astronomers. Determining their number density
in the form of their luminosity function is important to our
understanding of the Galaxy for a number of reasons.

Assuming a mass-magnitude relation, the subdwarf luminosity function
can be converted to a mass function, the form of which has key
implications for both Galactic and stellar astronomy. On a Galactic
level, the subdwarf luminosity function can be integrated to estimate
the number density of subdwarfs in the spheroid. With a large sample of
subdwarfs from a wide area, the number density as a function of line
of sight in the Galaxy can be used to accurately determine the
density law governing the spheroid. Additionally, although low
mass stars have been ruled out as significant dark matter candidates
\citep*{Bahcall et al.1994, Flynn et al.1996, Chabrier & Mera 1997,
Fields et al.1998}, the number density can place improved constraints
on their mass-to-light ratio, and can also be used to attain limits of
the dark matter contribution of brown dwarfs and white dwarfs through
extrapolation of the mass function.

For stellar astronomy, the mass function has important implications in
understanding the theories of stellar formation and
evolution. Comparing the mass function of old, metal-poor subdwarfs
with other types of stars allows determination of the mass function as
a function of epoch and metallicity, and hence about the conditions in
which the different populations of stars formed. For example, stars
forming in regions of higher temperature are expected to have higher
masses.

As well as providing the stellar mass function, the subdwarf
luminosity function can also enhance understanding of globular
cluster evolution. Comparing the luminosity and mass functions of
globular clusters with those of field stars of similar ages and
metallicities can highlight evolutionary processes in the
clusters. This method could verify whether the abnormally low
mass-to-light ratios of globulars is due to evaporation of their
low-mass members.

In this paper we first summarise the status of research into the
subdwarf luminosity function (Section~\ref{sec:recentresults}), before
describing our data (Section~\ref{sec:data}) and reduction methods
used (Sections~\ref{sec:methods_astrom}-\ref{sec:methods_lf}). Finally
(Section~\ref{sec:results}) we present our results and discuss their
consequences.


\section{Recent results} \label{sec:recentresults}
Spheroid members make up only a tiny proportion ($\sim$0.2\%) of the
stars in the solar neighbourhood. This factor, coupled with their
intrinsic faintness, makes subdwarfs hard to detect, and explains
why until very recently samples of known subdwarfs have been very
small.

There are two principal methods used to efficiently search for
subdwarfs. The traditional method is to exploit the high heliocentric
velocities of spheroid stars by imposing a minimum proper motion limit
on the sample. This biases selection towards spheroid members,
increasing the ratio of disc to spheroid stars from around 400:1 for a
volume-limited sample to about 4:1 for the proper motion-limited one.

The bias this selection introduces can be modelled with assumed
velocity ellipsoids for the disc and spheroid, although there remains a
significant sensitivity towards the kinematic models used. A further
disadvantage of this method arises with contamination of the sample
from high-velocity thick disc stars; however, imposing a strict lower
tangential velocity cut-off can render this negligible.

Most studies have employed the method of proper motion selection
to determine the spheroid luminosity function, simply because it
remains the most efficient method for obtaining samples of local
spheroid stars. Schmidt (1975) used it to provide the first estimate,
with improved determinations later offered by \citet{BC86}, and
many in the last decade (\citealt{D95}; \citealt{GR99};
\citealt{CR2000}; \citealt{Gould 2003}.)

The alternative method is to perform deep star counts, looking beyond
the solar neighbourhood and Galactic discs to probe pure regions of
spheroid stars \citep{GFB98}. This approach has been made possible
with the advent of sensitive telescopes such as the \emph{Hubble Space
Telescope} (HST), enabling accurate star-galaxy separation at large
distances where the stellar population contains only halo
members. Whilst samples derived in this way have none of the
contamination problems of the kinematic studies, there are
disadvantages. The great distances probed mean that trigonometric
parallaxes are currently unobtainable, so photometric parallaxes are
the only indicators of distance and hence luminosity. In addition,
there is no opportunity to obtain follow-up spectra of stars
identified in deep studies, and such samples can be more prone to bias
arising from unresolved binaries, although this is countered by the
high resolution of HST data in the \citet{GFB98} study.

Despite many investigations into the subdwarf luminosity function,
there remains significant disparity between its estimated shape and
normalisation. The deep star count luminosity function of \citet{GFB98}
predicts up to five times fewer stars than those of \citet{D95},
\citet{GR99} and \citet{Gould 2003} for M$_V \ga$ 8, and there are even
discrepancies among the kinematically-selected samples, with the
luminosity function of BC86 substantially lower than the others.

There are several possible explanations for the differences in the
luminosity functions. The first assumption must be that the effect is
due to systematics: incompleteness in samples (especially
\citealt{BC86}), inaccurate photometric parallaxes and differences in
kinematic models have all been postulated as contributing factors
(\citealt{GFB98}, \citealt{GR99}.)

However, the differences between the local, proper motion-selected
luminosity functions and the deep star count result of \citet{GFB98}
could partially be a manifestation of the spheroid density profile, as
recognised by \citet{GFB98} and \citet{GR99}. Several studies have
indicated that the spheroid has a highly flattened centre and a nearly
spherical outer component (\citealt{Sommer-Larsen & Zhen 1990,
CB2000}; \citealt*{Chen et al.2001} and references therein.) This
would result in a higher density of spheroid stars for the local
surveys since they would probe the flattened central ellipsoid while
the deeper study would sample regions beyond. However, it is unclear
whether this structure can completely account for the discrepancies
between the studies, and there are indications that the Galactic
stellar populations are more complex than those described by the usual
four discrete components of thin and thick discs, bulge and
halo. Uncertainties over these associated aspects of Galactic
structure and discrepancies in the shape and normalisation of the
subdwarf luminosity function motivate further investigation into its
true form.


\section{The data} \label{sec:data}

The sample in this study is drawn from overlapping regions of the
Sloan Digital Sky Survey (SDSS) Early Data Release (EDR)
\citep[][]{Stoughton et al.2002} and scans of POSS-I E plates in the
SuperCOSMOS Sky Survey (SSS; \citealt{Hambly et al.2001a, Hambly et
al.2001b, Hambly et al.2001c}.) The areas surveyed are the EDR South
and North Galactic Stripes (SGS \& NGS), outlined in
Table~\ref{tab:edr} and providing a total coverage of $\sim$394
deg$^2$. The SGS overlaps 12 SSS fields along the celestial equator,
and the NGS covers 16 plates (Table~\ref{tab:edr_fields}). Hereafter a
`field' refers to the area common to a POSS-I plate and one of the EDR
stripes: this forms a strip $\sim$2.5\degr wide along the celestial
equator at the centre of each plate.

The final $u^{'}g^{'}r^{'}i^{'}z^{'}$ SDSS filter system is described
by \citet{Fukugita et al.1996}. However, at the time of release of the
EDR there were uncertainties in the photometric calibration
\citep[see][]{Stoughton et al.2002}, so here we refer to the
preliminary photometry as $\sdssu\sdssg\sdssr\sdssi\sdssz$.

For the purposes of measuring proper motions we select the POSS-I
data for the first epoch ($\sim$1950) measures and the SDSS as the
second epoch ($\sim$1998). Whilst proper motions are available
from the SSS database we make this choice so as to maximise the
time baseline and hence the proper motion accuracy.

These data form an ideal sample from which to select high proper
motion spheroid stars: the epoch difference of $\sim$45 yr enables
proper motions to be determined accurately, whilst the SDSS photometry
provides precise magnitudes and colours.

\begin{table}
\caption{SDSS EDR regions} \label{tab:edr}
\begin{tabular}{l|c|c|c|} \hline
Region & RA & Dec & Area (deg$^2$)\\ \hline North Galactic Stripe
& 145\degr - 236\degr & 0\degr & 228 \\ \hline South Galactic
Stripe & 351\degr - 56\degr & 0\degr & 166 \\ \hline
\end{tabular}
\end{table}

\begin{table}
\caption{The survey fields in the EDR SGS \& NGS regions} 
\begin{tabular}{lccccll} \hline \label{tab:edr_fields}
POSS-I & $\alpha$ & $\delta$ & \ & \ & Plate & Area \\
Field      &  \multicolumn{2}{c}{(J2000.0)}  & l & b & Epoch  &
{($\times$10$^{-3}$ ster)}\\ \hline
\multicolumn{7}{l}{South Galactic Stripe} \\
0932 & 03 43 & +00 28 & 186 & -40 & 1954.0 & 2.66 \\
0363 & 03 19 & +00 31 & 181 & -45 & 1951.7 & 4.60 \\
1453 & 02 55 & +00 35 & 175 & -49 & 1955.8 & 4.61 \\
1283 & 02 31 & +00 38 & 168 & -53 & 1954.9 & 4.59 \\
0852 & 02 07 & +00 41 & 159 & -56 & 1953.8 & 4.61 \\
0362 & 01 43 & +00 44 & 149 & -59 & 1951.7 & 4.64 \\
1259 & 01 19 & +00 46 & 138 & -61 & 1954.8 & 4.61 \\
1196 & 00 55 & +00 47 & 125 & -62 & 1954.7 & 4.58 \\
0591 & 00 31 & +00 48 & 112 & -62 & 1952.7 & 4.59 \\
0319 & 00 07 & +00 48 & 100 & -60 & 1951.6 & 4.60 \\
0431 & 23 43 & +00 48 &  90 & -58 & 1951.9 & 4.64 \\
0834 & 23 19 & +00 48 &  80 & -54 & 1953.8 & 1.81 \\ \hline
\multicolumn{7}{l}{North Galactic Stripe} \\
0151 & 15 43 & -00 28 &   6 & 40 & 1950.5 & 1.57 \\
1402 & 15 19 & -00 32 &   1 & 45 & 1955.3 & 3.23 \\
1613 & 14 55 & -00 35 & 355 & 49 & 1957.3 & 3.56 \\
1440 & 14 31 & -00 39 & 348 & 53 & 1955.4 & 3.81 \\
1424 & 14 07 & -00 41 & 339 & 56 & 1955.4 & 4.06 \\
0465 & 13 43 & -00 44 & 329 & 59 & 1952.1 & 4.33 \\
1595 & 13 19 & -00 39 & 318 & 61 & 1956.3 & 4.39 \\
1578 & 12 55 & -00 47 & 305 & 62 & 1956.2 & 4.51 \\
1405 & 12 31 & -00 48 & 292 & 62 & 1955.3 & 4.59 \\
1401 & 12 07 & -00 48 & 280 & 60 & 1955.3 & 4.55 \\
0471 & 11 43 & -00 48 & 270 & 58 & 1952.1 & 4.50 \\
1400 & 11 19 & -00 48 & 261 & 54 & 1955.3 & 4.38 \\
1397 & 10 55 & -00 52 & 253 & 50 & 1955.3 & 4.27 \\
0467 & 10 31 & -00 45 & 247 & 46 & 1952.1 & 4.08 \\
0470 & 10 07 & -00 43 & 241 & 42 & 1952.1 & 3.81 \\
1318 & 09 43 & -00 40 & 237 & 37 & 1955.0 & 2.22 \\
\end{tabular}
\end{table}



\section{Methods: astrometry} \label{sec:methods_astrom}

The data are first paired, star-galaxy separated, and then
positional systematics are eliminated by means of an error mapping
algorithm. Proper motions are derived from the two epoch measures,
and quasars are used to verify the proper motion zero point and to
estimate the proper motion accuracy. A proper motion cut-off is
applied to avoid contamination from false detections and to
produce a clean high proper motion sample. The method of reduced
proper motion is then employed to select candidate subdwarfs and
the luminosity function is then derived from this sample.

\subsection{Pairing} \label{sec:pairing}

Objects common to both datasets are paired with an algorithm matching
on position, with each object paired to its nearest neighbour out to a
maximum radius of 10 arcsecs. With a mean epoch difference between the
POSS-I plates and the SDSS data of $\sim$45 yr, this corresponds to
an upper theoretical proper motion limit of $\sim$220 \masyr. Prior to
pairing the SDSS data are cut at r $\le$ 20.5, to avoid contamination
from spurious objects on the photographic plates. After these cuts
there are typically $\sim$50 000 paired objects per field.

\subsection{Star-galaxy separation} \label{sec:stargal}
The star-galaxy separation uses classifications given in the SDSS
catalogue. This has been shown to be at least 95\% accurate for r$^*
\le$ 21 \citep{Stoughton et al.2002}, adequate for our purposes and
better than could be obtained from analysis of the photographic
plates.

\subsection{Position-dependent astrometric errors} \label{sec:errmap}

The next stage is to remove position-dependent astrometric errors
which create well-known `swirl patterns' on Schmidt plates
\citep{Taff et al.1992} of systematic distortions between measured
plate positions and the expected tangent plane coordinates.

The SSS uses reference stars from the Tycho-2 catalogue \citep{Hog et
al.2000} to convert between the measured $(x,y)$ coordinates and the
celestial frame. The reference star catalogue positions are converted
to tangent plane (or standard) coordinates $(\xi, \eta)$, which are
then scaled for Schmidt cubic radial distortion \citep{Hambly et
al.2001c}. Large scale `swirl patterns' are then removed by applying
a mean distortion map, created by averaging the positional residuals
on a grid over a large number of plates from the appropriate survey. A
grid size of 10 arcminute is used since this is adequate to map the
expected $\sim$30 arcminute scale non-linear distortions. The mean
correction at each location on the plate is then obtained by
bilinearly interpolating in the grid. These corrections are added to
the standard coordinates of each star, the standard coordinates are
fitted to the measured positions, and the final conversion to
celestial coordinates achieved from the reference star plate solution.

The SSS data therefore has mean large-scale positional errors removed,
but there will still be plate-to-plate effects present as well as
systematic errors on smaller scales. These are dealt with by an error
mapping algorithm first used by \citet{EI95} (and employed for
deriving the SSS proper motions), which applies a large scale (10 arcmin or
1cm) correction to account for plate to plate systematics, and then a
further one-dimensional algorithm to remove small scale (2 arcmin)
errors. The procedure operates as follows:
\begin{enumerate}
\renewcommand{\theenumi}{(\arabic{enumi})}
\item A two-dimensional large-scale error mapper is applied by
dividing the field into an grid of 10 arcminute squares and measuring
the mean shift in the $x$ and $y$ coordinates between the two datasets
from all the stars in each cell. A 3x3 linear filter is used to reduce
the error function noise, and the $x$ and $y$ shift for each object in
the field is interpolated from the grid point corrections. The SDSS
position for each object (stars and galaxies) is then shifted with
respect to the SuperCOSMOS coordinates (the orientation of this shift
is immaterial) to remove large scale errors.  Stars and galaxies are
not treated separately at this stage so as to facilitate the proper
motion zero point shift by using the galaxies as described in
(\emph{iv}) below.

\item Small scale errors are then removed by a one-dimensional error
mapper, which calculates $x$ and $y$ errors as a function of both $x$
and $y$ by evaluating the mean shift for stars in each 2 arcminute
strip in both directions over the field. Median and then linear
filters are applied to smooth and reduce the noise in the functions,
which are then used to again shift the star and galaxy SDSS positions
with respect to the SuperCOSMOS measures. This process is iterated
until the mean shifts in each of the $x$ and $y$ strips is less than a
tolerance level (0.008 times the rms $x$ or $y$ shift.)

\item The two-dimensional error mapper is then applied again to
remove any large scale shifts introduced by the one-dimensional
algorithm.

\item At this point the stars have zero mean proper motion, whilst the
mean galaxy displacement is non-zero, since stars dominate the number
counts at these magnitudes. The galaxies are used to shift the proper
motion zero point by determining a least-squares fit of the galaxy
positions to find a six-coefficient linear plate model accounting for
zero-point, scale and orientation. A global translation is then
applied using this model to reset the galaxies to zero mean proper
motion, with the SDSS measures transformed by
\begin{eqnarray*}
x_{SDSS} & = & a + b x_{SSS} + c y_{SSS} \\
y_{SDSS} & = & d + e x_{SSS} + f y_{SSS} 
\end{eqnarray*}
where $a, b, c, d, e$ and $f$ are the plate coefficients determined
from the least-squares fit to the galaxy positions.
\end{enumerate}

\begin{figure}
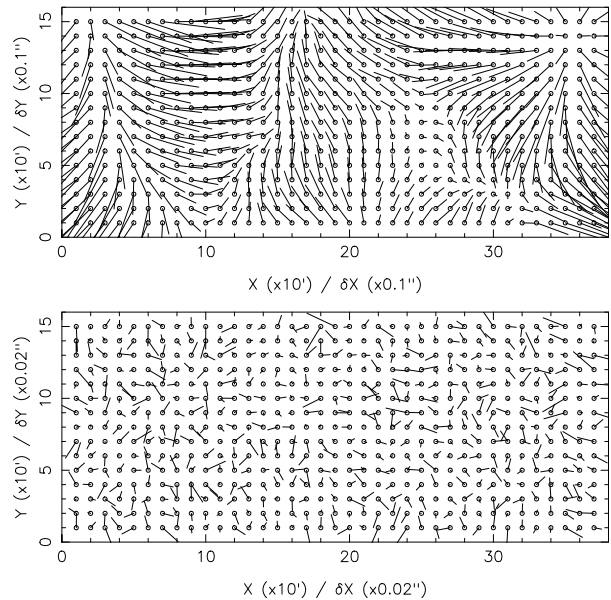

\centering
\includegraphics[height=90mm,angle=270,trim=70 0 0 0 ]{fig01a.eps}
\includegraphics[height=90mm,angle=270,trim=190 0 70 0 ]{fig01b.eps}
\caption{Before and after the error mapper is applied in field 0363 in
the SGS, which covers a 2\fdg5-high strip along the centre of the
Schmidt plate. The `lollipops' show the relative size of the binned,
smoothed and filtered positional errors at each point in the
field. Note that the binning and smoothing means that these plots show
systematic, not random, errors and that the scale of the errors in the
bottom figure is five times smaller than that in the top figure.}
\label{fig:errmap}
\end{figure}

The results of applying this error mapper can be seen in Figure
\ref{fig:errmap}, which shows a typical survey field covering a strip
along the centre of the Schmidt plate that is 6\fdg25 wide and
2\fdg5 high. The first panel shows the field after the SSS mean
large-scale distortion algorithm has been used, but before the error
mapping algorithm described above has been applied, with the large
scale `swirl pattern' of the photographic data still evident. After
the error mapper there are only very small residual systematic errors
in position, and the remaining random errors have an rms of only
$\sim$0.3 arcsec.

\subsection{Photometric-dependent astrometric errors}
\begin{figure}
\centering
\includegraphics[height=85mm, angle=270,trim=0 60 0 20 ]{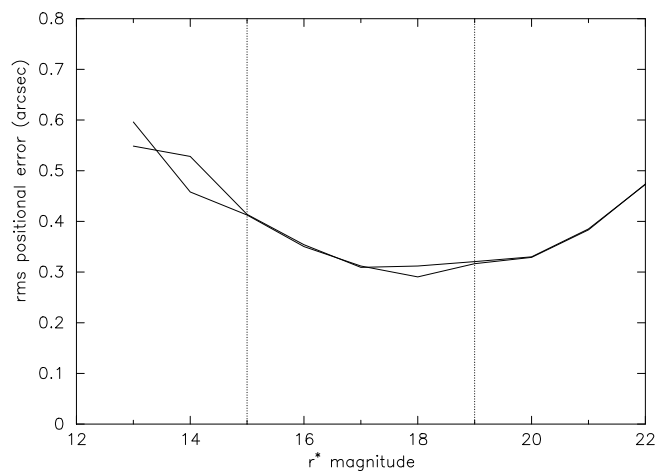}
\caption{The rms of the random positional errors in $x$ and $y$ as a
function of magnitude for a typical field in the NGS, once the error
mapping algorithm has been applied. In the magnitude range of interest
(dashed lines) there is only variation with magnitude at the
$\sim0.1$ arcsec level, and the errors are smallest where the majority
of the subdwarf sample is found ($\sdssr \approx 18$). These
considerations indicate that systematic magnitude effects will not
significantly affect the derived astrometric parameters.}
\label{fig:posmagerr}
\end{figure}

Systematic biases dependent on the magnitude of targets are
particularly prevalent in photographic astronomy, and could affect the
selection or derived astrometric parameters of stars. These systematic
magnitude errors cannot be corrected for in the data, since this could
remove real effects present, such as brighter stars tending to have
higher proper motions due to their mean closer proximity to the
Sun. However, as shown in Figure~\ref{fig:posmagerr}, the influence of
these errors is likely to be small for the high proper motion sample:
their size is only $\la$8 \masyr\ (with a $\sim$45-yr epoch
difference) for a sample selected with a proper motion cut five times
larger (see Section~\ref{sec:pmselection}), and they have a systematic
variation with magnitude only at the $\sim$0.1 arcsec level ($\sim$2
\masyr). Additionally, this systematic error applies to the error
difference with magnitude across the whole $15 \la \sdssr \la 19.5$
range, whereas the subdwarf sample is dominated by stars with $\sdssr
\ga17$ \citep{Chen et al.2001, Gould 2003}. The more pronounced errors
at $\sdssr \la 16$ seen in Figure~\ref{fig:posmagerr} are therefore
unlikely to have a large effect on the derived proper motions in this
study, and any magnitude-dependent errors should not lead to
significant astrometric or selection effects.

Similarly, systematic colour errors are expected to have little effect
on the astrometry of the sample. Biases arising from variations in
atmospheric effects are unlikely since all observations were taken on
or near the meridian, and the telescopes used in the survey are at
similar latitudes: Apache Point Observatory where the SDSS
observations were made is at 32\degr N 46', and the Palomar
Observatory is at 33\degr N 21'. Discrepancies in positions between
the SDSS and SSS data due to differential colour refraction will also
contribute only a small effect, since the SDSS $\sdssr$ and POSS-I E
passbands are very similar (centred at $\lambda\sim$6400\AA), and only
\emph{relative} refraction differences are important when determining
proper motions.

\subsection{Proper motion accuracy} \label{sec:pmaccuracy}
\subsubsection{Quasar motions}

The proper motions of all the stars in each field are measured after
error mapping, and an external check on the accuracy of these is
obtained from analysis of the quasars in each field.  The Veron and
Veron 10th QSO catalogue \citep{Veron2001} and the SDSS QSO catalogue
\citep{EDRQSO} provide a mean of 135 and 172 quasars per field in our
data for the SGS and NGS respectively, which are used to estimate the
proper motion accuracy.

\begin{figure}
\centering
\includegraphics[height=100mm,trim=10 0 0 0]{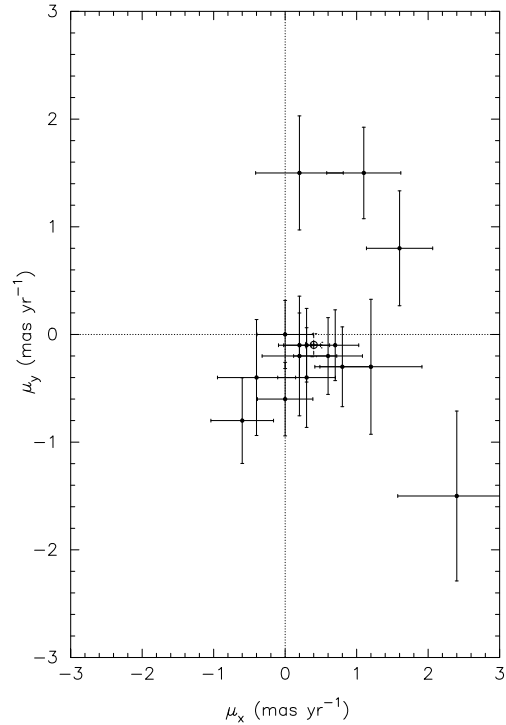}
\caption{Proper motion zero points as measured from quasars for the
NGS fields (points) and the combined NGS dataset (open circle).}
\label{fig:qso_zero}
\end{figure}

The mean spatial discrepancies between the quasars' SDSS and
SuperCOSMOS coordinates are divided by the epoch difference to
estimate the rms proper motion error for each field, under the
assumption that the quasars should have zero motion. Due to variations
in the quality of the plate data this error deviates significantly
from field to field, but is $\sim$5.0 - 9.0 \masyr\ for the SGS and
$\sim$5.0 - 10.0 \masyr\ for the NGS. However, it should be noted that
the true proper motion errors of the proper motion sample are liable
to be less than these values, since the quasars are fainter and have
poorer centroiding than the majority of the subdwarfs in this
study. The zero point of the quasar proper motions is consistent with
zero, as shown in Figure~\ref{fig:qso_zero}.

\subsubsection{Comparison with SSS proper motions}

The accuracy of the derived proper motions can also be ascertained by
comparison with those published in the SuperCOSMOS Sky Survey
\citep{Hambly et al.2001c}. These are measured from the B$_{\rm J}$
and R plates of the survey, with systematic positional errors removed
as in this study (Section~\ref{sec:errmap}) and differential colour
refraction effects also accounted for. The SSS proper motions are
currently derived from data at two epochs, with a median epoch
difference between the B$_{\rm J}$ and R plates of 15 yr, although
future releases will soon utilise measures from four plates where
available.

To compare the proper motions derived here with the SSS results we
have analysed SSS UKJ/UKR field 866, which overlaps the POSS-I fields
1440 and 1424 in the NGS. The B$_{\rm J}$ and R plates in field 866
have an epoch difference of 17 yr, so with good plate quality this
comparison uses SSS proper motions that are likely to be of above
average accuracy.

\begin{figure}
\centering
\includegraphics[height=80mm,trim=0 0 0 0]{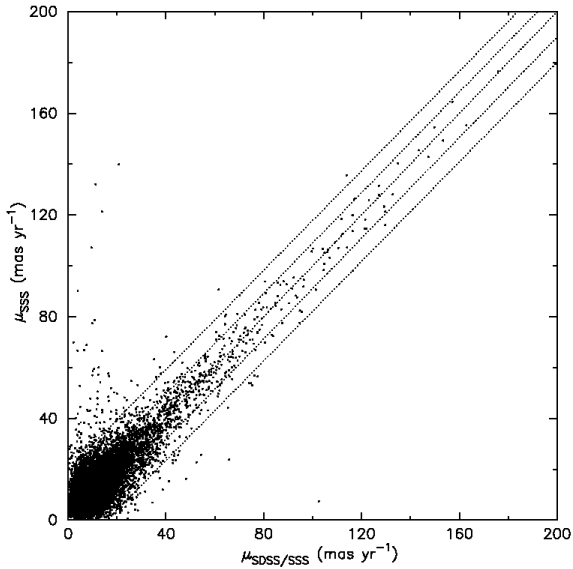}
\caption{Comparison of the proper motions derived in this study
($\mu_{\rm SDSS/SSS}$) with those from the SuperCOSMOS Sky Survey
($\mu_{\rm SSS}$) in NGS fields 1424 and 1440. Some $\sim$99\% of
stars fall within the 2$\sigma$ boundaries of a perfect correlation,
and the linear correlation coefficient is 0.85. The lack of stars
along the $\mu_{\rm SDSS/SSS}$ axis shows close agreement between the
SDSS/SSS measures and the corresponding SSS estimates for stars with
high SDSS/SSS proper motions, demonstrating the good reliability and
negligible contamination of our high proper motion sample.}
\label{fig:comparepm}
\end{figure}

The stars in our sample are paired with the SSS data, with image
quality criteria (such as restrictions on blended objects, see
Section~\ref{sec:ImageQual}) applied as in our analysis. Our proper motions
derived from the SDSS/SSS data are then compared with the SSS proper
motions for each star. Figure~\ref{fig:comparepm} shows this
comparison for stars with $15.0 \la \sdssr \la 19.5$, along with
1$\sigma$ and 2$\sigma$ deviations from a perfect correlation,
assuming standard deviations of 10 \masyr\ and 8 \masyr\ for the SSS
and SDSS/SSS proper motions respectively.  There is good correlation
between the measures, with a linear correlation coefficient of
$\sim$0.85 and $\sim$99\% of the stars falling with the 2$\sigma$
error bars, indicating negligible systematic effects. Note that there
are virtually no stars that have discrepantly high SDSS/SSS proper
motions compared to the SSS measures, although there are some for
which the SSS proper motion estimate appears to be significantly too
high. This is likely to be due to the facts that the SSS results are
derived from a smaller time baseline and from photographic material at
both epochs, increasing the likelihood of objects with spurious proper
motions entering the SSS sample. The excellent consistency of the
SDSS/SSS proper motions compared to the SSS measures for stars with
high SDSS/SSS proper motions demonstrates the good reliability of our
strict high proper motion sample selection and suggests negligible
contamination from objects with false motions.

\subsection{The high proper motion sample}
\subsubsection{Image quality criteria} \label{sec:ImageQual}

Prior to the proper motion cut being applied, the data are subject to
criteria to ensure that only objects with stellar and high quality
images are included in the final sample. From the SSS data the
following characteristics of each image are restricted (see
\citealt{Hambly et al.2001a} for more details):

\begin{itemize}{}
\item \emph{Blend:} Objects appearing as blended in the SSS database
are rejected.

\item \emph{Profile Classification Statistic, $\eta$:} This quantifies
the `stellarness' of an object by comparing the residuals of the areal
profile of an image with that of an average stellar template. Objects
with $\eta \ge 4 \sigma$ are rejected.

\item \emph{Quality:} During processing the SSS data are assigned a quality
flag, which is affected by circumstances such as an image being very
large, bright, fragmented or close to a bright star or plate
boundary. Images with a quality value of greater than 127 are
rejected.

\item \emph{Ellipticity:} The ellipticity of an image is calculated from the
weighted semi-minor and semi-major axes given by the SSS
processing. Only objects with e $\le$ 0.25 are included in the
sample.
\end{itemize}

\subsubsection{Proper motion selection} \label{sec:pmselection}

A lower proper motion limit is applied to the sample of paired stars
for two principal reasons: to amplify the proportion of spheroid stars
selected and to create a `cleaner' proper motion sample. The proper
motion selection magnifies the contribution from the higher-velocity
spheroid population, since they are effectively sampled over larger
volumes than the lower-velocity disc stars \citep{Reid 1997,
CR2000}. The number of stars of each population in a proper motion
selected sample is proportional to the mean population tangential
velocity:
\be \label{eq:mu_amp1} N(\mu > \mu_{min}) \propto \rho_0 \left< V_T
\right> ^3, \ee with $\rho_0$ the local space density of the
population \citep*{Hanson 1983, Reid 1984}. This therefore amplifies
the contribution of the higher velocity population above the ratio of
the local space densities by the amount:
\be\label{eq:mu_amp2}
A_\mu = \left(\frac{\left< V_T^1 \right>}{\left< V_T^2 \right>}\right)^3.
\ee
This amplification has a dramatic effect on the likelihood of high
velocity stars entering the proper motion sample, and demonstrates the
efficiency of proper motion selection in selecting spheroid stars. As seen
in \citet{CR2000}, a spheroid to disc number ratio of $N_{disc}$:$N_{spheroid}$
= 400:1 for a volume-limited sample can be increased to
$N_{disc}$:$N_{spheroid}$ = 5:1 for a proper motion sample.

The second effect of applying a minimum proper motion limit to the
sample is to include stars with small relative errors in proper
motion, avoiding those with marginal proper motions arising from
errors in the positions or pairing. This results in a `cleaner'
reduced proper motion diagram (Section~\ref{sec:RPM}), and a more accurate
subdwarf selection.

The lower proper motion limit can be applied either on a global basis,
with one limit for all fields, or for each field individually, using
the proper motion error estimates from the quasar positions in each
field (Section~\ref{sec:pmaccuracy}). Initially this latter approach
was used, but evidence of systematic differences between the proper
motion accuracies derived for the SGS and NGS led to the adoption of a
global proper motion minimum for the entire SGS and NGS sample. The
rms of the quasar `proper motions' in each stripe was found to be
$\sigma_\mu \approx 6.7$ \masyr\ for the SGS and $\sigma_\mu \approx
8.1$ \masyr\ for the NGS. In order to avoid the low proper motion
`tail' in each stripe the more conservative value of the NGS error was
used for all fields. A 5$\sigma_{\mu}$ cut on the proper motions was
applied in defining the lower limit, so that the value used for the
entire sample was:
\be \label{eq:mu_min} \rm \mu_{min} = 40.5\ mas\ yr^{-1}. \ee
A maximum proper motion limit is also adopted so as to ensure sample
completeness. This limit is theoretically determined by the maximum
pairing radius of 10 arcsec, corresponding to $\rm{\mu_{max} \approx
220\ mas\ yr^{-1}}$. However, the actual maximum is somewhat smaller
than this value due to systematic shifts in position affecting the
pairing process; we therefore adopt a conservative $\rm{\mu_{max} =
160\ mas\ yr^{-1}}$ from proper motion number counts (see
Section~\ref{sec:samplerr}.)

\begin{figure}
\centering
\includegraphics[height=100mm,trim=10 0 0 0]{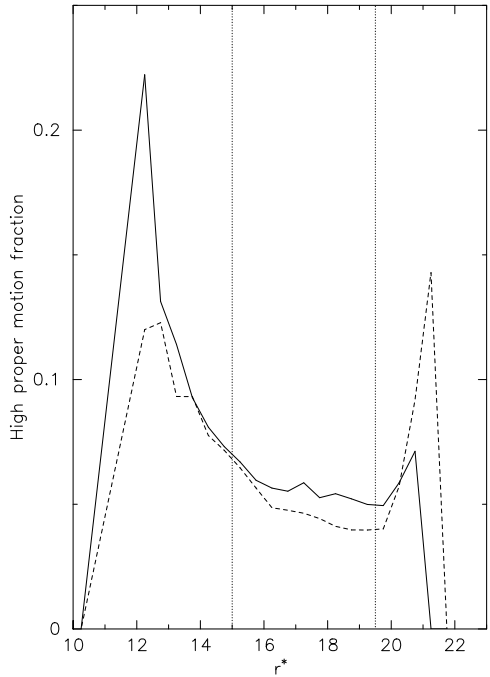}
\caption{The numbers of high proper motion stars in the SGS (solid
line) and NGS (dashed line), expressed as a fraction of the total
number of paired stars and plotted as a function of magnitude. For the
magnitude range of the sample (between the dotted lines) the
proportion in each stripe agree to within a few percent, indicating no
systematic errors in the derivation of proper motions or high proper
motion selection.}
\label{fig:hpmfrac}
\end{figure}

Figure~\ref{fig:hpmfrac} shows the relative proportions of high proper
motion stars with $\mu_{min} \la \mu \la \mu_{max}$ in the SGS and
NGS. Although there are discrepancies at the bright and faint ends,
over the magnitude range of the sample selected (15 $\la \sdssr \la$
19) the fraction of stars passing the proper motion criteria agree to
within a few percent.

\section{Methods: the subdwarf sample} \label{sec:methods_sample}
\subsection{Reduced proper motion} \label{sec:RPM}

Although applying a lower proper motion limit produces a cleaner
sample of fast-moving stars, this will consist not only of subdwarfs
but also white dwarfs and high velocity members of the disc. In order
to identify the candidate subdwarfs, the \emph{reduced proper motion}
of each star is used as a discriminator. Defined as 
\be H = m +
5\log\mu + 5 = M + 5\log V_T - 3.379, \label{eq:rpm} 
\ee 
this parameter separates effectively high-velocity (V$_T$), low-luminosity
(M) subdwarfs from the white dwarf and thin disc populations. This is
seen in the reduced proper motion diagram (RPMD) of H$_{r^*}$ plotted
against $(\sdssri)$: Figures \ref{fig:rpmd_sgs} and \ref{fig:rpmd_ngs}
show the RPMDs for all stars with $\mu > \mu_{min}$ in the SGS and NGS
respectively. The fainter and higher tangential velocity subdwarfs
form a distinct sequence below the thin disc main sequence, whilst the
white dwarfs populate the bluest part of the diagram.

The colour index $(\sdssri)$ was chosen because of its effectiveness
at separating subdwarfs from higher-metallicity stars on
colour-magnitude diagrams, due to their lower opacities and higher
temperatures at a given mass and luminosity resulting in a larger
proportion of their flux being emitted at optical wavelengths
\citep{Gizis 1997, GR99}. The $(\sdssri)$ index is a good
temperature/spectral type indicator for the M dwarfs which dominate
this survey (see Section~\ref{sec:est_metal}; \citealt{Lenz et al.1998,
Hawley et al.2002}), so on a reduced proper motion or colour-magnitude
diagram it enables clear distinction between the solar-metallicity
stars and the hotter (bluer) subdwarfs at each absolute magnitude. The
$(\sdssri)$ colour is similar to $(R-I_C)$, proven to be effective at
separating subdwarfs in this manner, and covers the TiO and CaH
spectral features important for subdwarf classification \citep{Gizis
1997, GR99}. We investigated other appropriate SDSS indices for use in
the RPMD, but $(\sdssri)$ proved to be the most effective for subdwarf
identification.

\subsection{Reddening and extinction}
With all the fields at high Galactic latitudes ($|b| \ge 37 \degr$),
interstellar extinction and reddening are unlikely to have a
significant impact on this study. To confirm this we use the
high-resolution COBE/DIRBE dust maps of \citet{Schlegel et al.1998} to
estimate the reddening and extinction at each field centre, using the
given wavelength coefficients to convert $E_{B-V}$ and $A_V$ to
$E_{\sdssri}$ and $A_{\sdssr}$. We find that the majority (23 out of
28) of the SGS and NGS fields have $E_{\sdssri}<0.04$, and that the
overall mean is $E_{\sdssri}=0.03$. The maximum reddening is
$E_{\sdssri}=0.08$ (for field 0932), and only three fields have
$E_{\sdssri} \ge 0.05$, two of which are narrow fields at the end of a
stripe and hence contain only a small fraction of the total sample. In
applying these estimates we assume that there is no differential
reddening across the fields, and that all obscuration is in the
foreground. This assumption is valid for our study, since the
reddening layer scale height is $\sim$100 pc \citep{Chen et al.1999},
and our survey is sensitive to very few stars within this distance
(see Section~\ref{sec:photparx}.)

\begin{figure}
\centering
\includegraphics[height=90mm,trim=20 10 20 0]{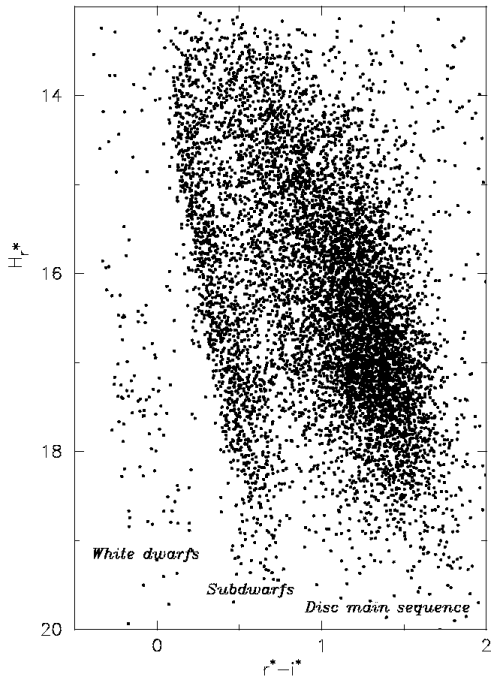}
\caption{The RPMD for stars with $\mu \ge \mu_{\rm{min}}$ in the
SGS. The three separate sequences of white dwarfs, subdwarfs and thin
disc stars are clear.}
\label{fig:rpmd_sgs}
\end{figure}

\begin{figure}
\centering
\includegraphics[height=90mm,trim=20 10 20 0]{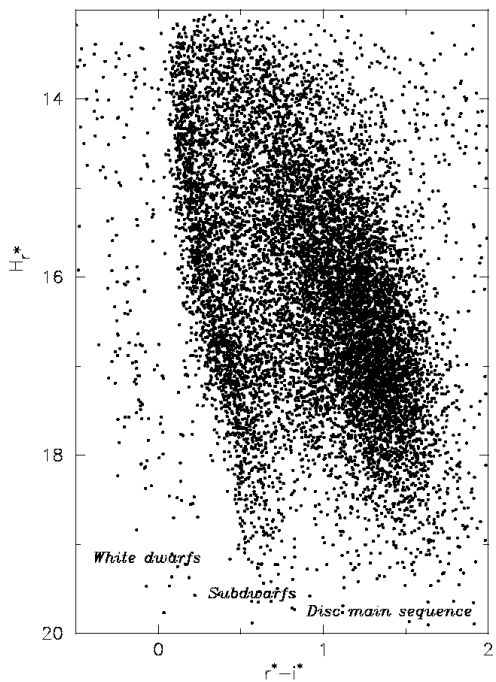}
\caption{The RPMD for stars with $\mu \ge \mu_{\rm{min}}$ in the NGS.}
\label{fig:rpmd_ngs}
\end{figure}

With photometric accuracy of $\sim$3\% in the $\sdssr$ and $\sdssi$
bands of the EDR \citep{Stoughton et al.2002}, the size of the
reddening effect is comparable to that of the $\sigma_{\sdssri}\approx
0.04$ colour errors. Inspection of the reduced proper motion diagrams
(Figures~\ref{fig:rpmd_sgs} and \ref{fig:rpmd_ngs}) shows that
adjustments of this order would have little effect on the
``tightness'' of the population locii and subsequent subdwarf
selection. The mean extinction over all fields is $A_r = 0.126$, which
if not corrected for would result in derived photometric parallaxes
being underestimated by up to $\sim$5\%, less than the expected
distance errors. These considerations suggest that the effects of
reddening and extinction on the study are negligible, and so
corrections are not applied to the data.

\subsection{Tangential velocity cut-off} \label{sec:vtcutoff}
Although the RPMD separates the subdwarfs from the thin disc stars and
white dwarfs, there is some overlap on the diagram with the locus of
thick disc stars. These have relatively high velocities, and hence are
responsible for the objects lying between the old disc and subdwarf
sequences. Thick disc stars have a much higher local number density
than subdwarfs, and the inclusion of even a tiny proportion of this
population in the sample can result in a significant overestimation of
the spheroid density \citep{BC86}.

\begin{figure}
\includegraphics[width=55mm,angle=270,trim=10 20 0 0]{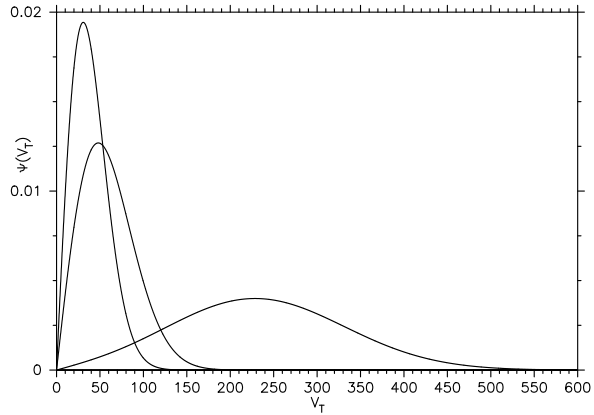}
\caption{Simulated tangential velocity distributions for the thin and
thick disc and spheroid (in order of increasing velocity) in the
direction of field 0363 in the SGS}
\label{fig:vtdist}
\end{figure}

However, the contamination of the sample by these stars can be avoided
by the introduction of a cut-off in tangential velocity.  Figure
\ref{fig:vtdist} shows calculated tangential velocity distributions of
the disc and spheroid populations in the direction of one of the SGS
fields, assuming a velocity ellipsoid described by a solar motion and
Gaussian dispersions (Table \ref{tab:kinematics}) and following the
method given in \citet[][ p285]{Murray 1983}.

For our assumed velocity ellipsoids we adopt the disc kinematics
derived from two analyses of local M dwarfs \citep*{Reid et al.1995,
Hawley et al.1996}, and the spheroid ellipsoid derived from the study of a
large sample of low metallicity stars by \citet{CB2000}. The
parameters of these ellipsoids are given in Table
\ref{tab:kinematics}.

It is clear that a limit of $V_T >$ 200 \kms\ will exclude all
but a negligible proportion of the thick disc population: our
calculations suggest that a maximum of just 0.04\% of the thick disc
population will be included in such a sample.  This cut-off will also
cause the low-velocity tail of the spheroid population to be excluded from
the sample, but our calculation results allow us to correspondingly correct
the derived luminosity functions (see Section~\ref{sec:discfrac}.)

\begin{table}
\caption{Adopted velocity ellipsoids. The assumed disc (\citealt{Reid
et al.1995}; \citealt{Hawley et al.1996}) and spheroid \citep{CB2000}
stellar kinematics used in the tangential velocity calculations. All
speeds are in kms$^{-1}$, and U, V, W denote the usual Galactic
coordinate triad in the respective directions of the Galactic center,
direction of rotation and NGP.}
\label{tab:kinematics}
\centering
\begin{tabular}{lcccccc} \hline
Population & U$_{\sun}$ & V$_{\sun}$ & W$_{\sun}$ & $\sigma_U$ & $\sigma_V$ & $\sigma_W$ \\ \hline
Spheroid & -26 & -199 & -12 & 141 & 106 & 94\\
Thin Disc & 10 & -5 & -7 & 41 & 27 & 21 \\
Thick Disc & 10 & -23 & -7 & 52 & 45 & 32 \\ \hline
\end{tabular}
\end{table}

\subsection{Selecting candidate subdwarfs}

\subsubsection{The colour-magnitude relations}\label{sec:cmr}

Equation (\ref{eq:rpm}) suggests that with an assumed colour-magnitude
relation, lines of constant $V_T$ can be plotted on the RPMD. A
spheroid-disc separating line of $V_T$ = 200 \kms\ can then be applied
to the RPMD in order to select candidate subdwarfs, with an upper RPM
limit used to exclude white dwarfs from the sample. An accurate colour
magnitude relation is also required to derive photometric parallaxes
for each star in the sample, necessary for computation of the
luminosity function.

Obtaining a reliable colour-magnitude relation for subdwarf stars
empirically is problematic, due to the existence of relatively few
known subdwarfs with accurate trigonometric parallaxes. Theoretical
relations derived from model atmospheres exist as an alternative
\citep{BCAH97}, and these provide excellent matches to cluster
sequences and closely follow the majority of field star data,
especially in the infrared \citep{BCAH97, BCAH98, Chabrier
2003}. However, the metal-rich versions ([m/H] $\ga -1.0$) of these
models suffer from discrepancies with the lower end of the observed
main sequence ($V-I_C \ga 2.0$) in optical colours \citep{BCAH98,
Chabrier 2003}. Whilst lower-metallicity models are expected to be
less affected in this way, when matched to field stars they still
yield poor reproductions of the observed colours at redder wavelengths
\citep{BCAH97, Chabrier 2003}, and exhibit possible systematic errors
in metallicity when used to estimate abundances from colour-magnitude
diagrams \citep{Gizis 1997, GFB98}. These factors could significantly
affect the ability of the models to provide the reliable optical
colour-magnitude relations required here, calculating an expected
absolute magnitude for each subdwarf given an observed colour and
metallicity estimate. Therefore, whilst the latest theoretical model
atmospheres provide good matches to most of the stellar physics and
observed parameters, the possibility of systematic colour offsets,
perhaps wavelength-dependent, leads us to choose to employ an
empirical colour-magnitude relation for this study.

An obstacle in applying a colour-magnitude relation to halo stars is
the large range of metallicities present in the spheroid, and hence
the spread of absolute magnitude for a given colour. Without spectra
to obtain metallicities we deal with this degeneracy in two ways: by
using one approximate colour magnitude relation for selecting
subdwarfs on the RPMD, and a more accurate estimate which predicts
metallicity from observed colours for deriving photometric parallaxes.

\subsubsection{The colour-magnitude relation for selecting subdwarfs} \label{sec:cmrsel}

The colour-magnitude relation used to select subdwarfs from the RPMD
is obtained in a similar way to that of \citet{GR99}. We derive
separate relations for low and high metallicity subdwarfs, following
the classifications defined by \citet{Gizis 1997} of stars with [Fe/H]
$\approx -1.2 \pm$0.3 as subwarfs (sd), and with [Fe/H] $\approx
-2.0\pm$0.5 as extreme subwarfs (esd).

To derive a colour-magnitude relation we use calibrating subdwarfs
from the compilations of \citet{Gizis 1997} and \citet*{Reid et
al.2001}. Both of these studies use the spectral classification scheme
described in \citet{Gizis 1997} to identify subdwarfs, and give
accurate trigonometric parallaxes for their samples. These are derived
principally from the USNO \citep*{Monet et al.1992} and Yale (4th ed;
\citealt*{van Altena et al.1995}) catalogues in the case of Gizis, and
from the Hipparcos catalogue \citep*{Hipp} in the case of
\citet{Reid et al.2001}.

\begin{figure}
\includegraphics[width=60mm,angle=270,trim=0 20 0 20]{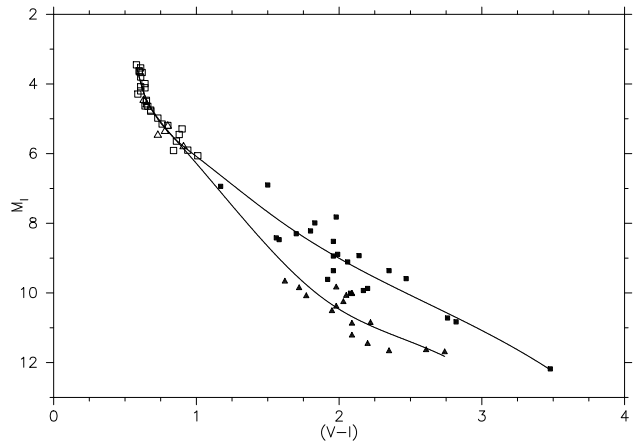}
\caption{Colour-magnitude relations derived from subdwarfs with
accurate trigonometric parallaxes. Squares denote subdwarfs and
triangles extreme subdwarfs; filled symbols are subdwarfs from
\citet{Gizis 1997} and open symbols are those from \citet{Reid et
al.2001}}
\label{fig:parxsd}
\end{figure}

We consider only subdwarfs with $\sigma_{\pi}/\pi < $0.2 so as to
ensure accuracy of the absolute magnitudes, producing a total of 35
subdwarfs and extreme subdwarfs from Gizis, and 38 from Reid et al. We
plot these stars on a M$_I$, $(V-\Ic$) Hertzsprung-Russell diagram
(Johnson $V$ and Cousins $\Ic$), and fit a colour-magnitude relation
to the sdM and esdM separately. Stellar colour-magnitude relations are
poorly matched by a linear relation due to an inflection in the
subdwarf main sequence at V-$\Ic \sim$1.5 \citep{BCAH97},
so we match a cubic spline to the parallax data (Figure
\ref{fig:parxsd}.)

There is currently no published SDSS photometry for stars with
accurate trigonometric parallaxes, so it is necessary to first
derive the colour-magnitude relation in the standard
Johnson/Cousins photometric systems, and then convert the colours
and magnitudes to SDSS $\sdssr$ and $\sdssri$. This is done by
means of the relation

\be \label{eq:cmrconv} M_{r^*} = r^*-V + (V-\Ic) + M_I, \ee
combined with two-colour relations to convert from SDSS
($\sdssri$) to ($\sdssr-V$) and ($V-\Ic$).

\begin{figure}
\centering
\includegraphics[width=65mm,trim=0 0 0 0]{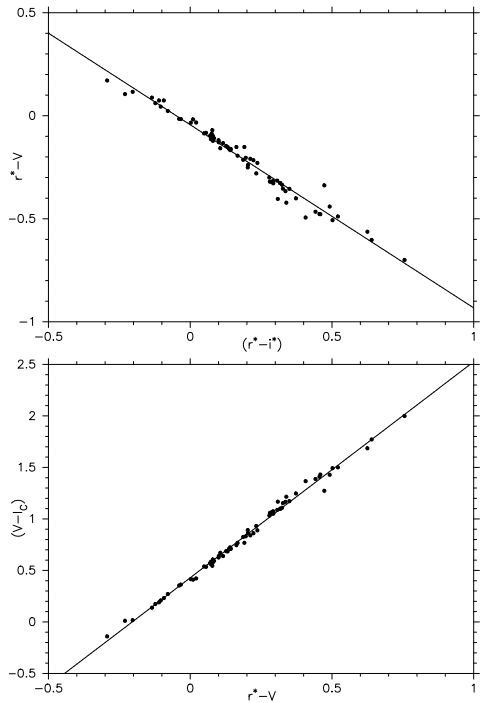}
\caption{Two-colour relations from the SDSS to Johnson/Cousins
photometric systems for disc stars in the SDSS standard star
catalogue. There is little to suggest that these relations are
strongly dependent on metallicity.} \label{fig:twocol}
\end{figure}

These two-colour relations should ideally be obtained from subdwarfs
with accurate photometry and similar metallicities to those in our
sample. However, of the subdwarfs in the SDSS Standard Star Catalogue
\citep{Smith et al.2002}, very few have published $V\Ic$
photometry and span only a small range in colour.  We therefore fit
linear two-colour relations between the Johnson/Cousins and SDSS
systems for higher-metallicity disc stars instead. Despite this,
applying these fits to the few subdwarf stars with SDSS and $V\Ic$ magnitudes suggests that the disc metallicity fits apply well to
the spheroid metallicity stars. The relations derived are 

\be
\label{eq:twocol1} \sdssr -V =-0.889(\sdssri)-0.040, \ee \be
\label{eq:twocol2} V-\Ic = 2.097(\sdssri)+0.429, \ee and are shown in Figure
\ref{fig:twocol}.

\subsubsection{Selecting candidate subdwarfs}\label{sec:selectsd}

With an approximate colour-magnitude relation with which to estimate
M$_{r^*}$ given ($\sdssri$) for an assumed subdwarf, we can make use
of Equation~(\ref{eq:rpm}) to plot lines of constant $V_{\rm T}$ on the
RPMD.

\begin{figure}
\centering 
\includegraphics[width=65mm,trim=0 10 0 30]{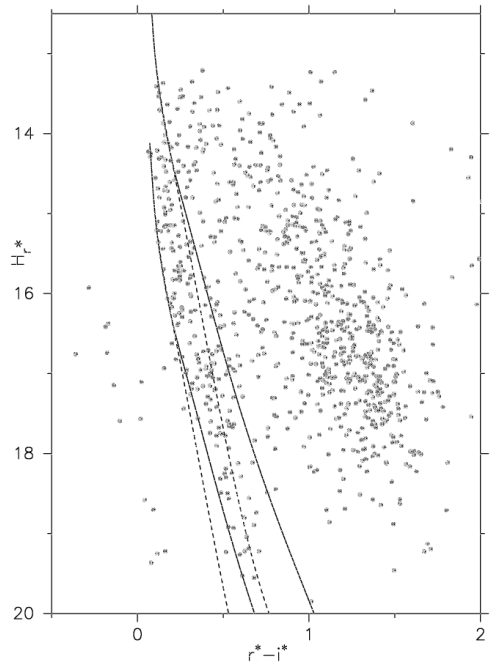}
\caption{The RPMD for high proper motion stars in NGS field 0465, with
the subdwarf (solid lines) and extreme subdwarf (dashed lines)
selection lines corresponding to V$_{\rm T}$ = 200 and 500 kms$^{-1}$
overlaid.}
\label{fig:rpmdvt}
\end{figure}

Figure~\ref{fig:rpmdvt} shows the RPMD for the high proper motion
stars in one of our fields (POSS-I field 0465 in the NGS), with the
lines corresponding to 200 and 500 \kms\ that we use to identify our
candidate subdwarfs. These limits clearly provide a good match to the
observed subdwarf sequence, and are applied on a field-by-field basis
to account for the varying kinematics between fields.  The lower
velocity bound is applied to avoid contamination from thick disc
stars, whereas the upper limit is applied to prevent selection of
white dwarfs. With knowledge of the expected tangential velocity
distributions for the disc and spheroid stellar populations (Section~
\ref{sec:vtcutoff}) we can estimate the fraction of spheroid and thick
disc stars that are expected to fall within this velocity range to
compensate for selection effects (Section~\ref{sec:discfrac}). The numbers
of high proper motion stars and subdwarfs passing all selection
criteria (including metallicity constraints for subdwarfs) are shown
in Table \ref{tab:nstar}.

\begin{table}
\caption{The numbers of high proper motion stars $(N_{hpm})$ and
subdwarfs $(N_{sd})$ passing all magnitude, proper motion, quality and
metallicity selection criteria for each field.}
\label{tab:nstar}
\centering
SGS \hspace*{35mm} NGS \\
\begin{tabular}[t]{lcccccc} \hline
Field & $N_{hpm}$ & $N_{sd}$  \\ \hline
0932& 296&  15 \\ 
0363& 757&  68 \\ 
1453& 703&  81 \\ 
1283& 747&  85 \\ 
0852& 709&  80 \\ 
0362& 814&  91 \\ 
1259& 804&  89 \\ 
1196& 790& 107 \\ 
0591& 875&  96 \\ 
0319& 864&  84 \\ 
0431& 847&  90 \\ 
0834& 320&  32 \\ \hline
Total:&  8526&  918 \\ \hline
\end{tabular}
\begin{tabular}[t]{lcccccc} \hline
Field & $N_{hpm}$ & $N_{sd}$  \\ \hline
0151& 256&  44 \\ 
1402& 703& 143 \\ 
1613& 700& 125 \\ 
1440& 814& 145 \\ 
1424& 866& 145 \\ 
0465& 910& 164 \\ 
1595& 853& 130 \\ 
1578& 802& 128 \\ 
1405& 947& 125 \\ 
1401& 1005& 125 \\ 
0471& 947& 111 \\ 
1400& 813&  89 \\ 
1397& 704&  69 \\ 
0467& 762&  76 \\ 
0470& 600&  56 \\ 
1318& 282&  21 \\ \hline
Total:& 11964& 1696 \\ \hline
\end{tabular}
\end{table}

\subsection{Metallicities and photometric parallaxes}
\subsubsection{Estimating metallicities} \label{sec:est_metal}

The division of the subdwarf colour-magnitude relation into two
metallicity ranges is clearly a poor approximation, and one that is
unsuitable for use in obtaining photometric parallaxes of the
candidate stars. The spread of nearly two magnitudes in luminosity
between metal-rich and metal-poor subdwarfs at $(V-\Ic) \approx$ 2 could
lead to significant errors in the distance estimates, so a more robust
method which accounts for the absolute magnitude-metallicity
degeneracy is required.

\begin{figure}
\centering \includegraphics[width=75mm,trim=0 10 0 30]{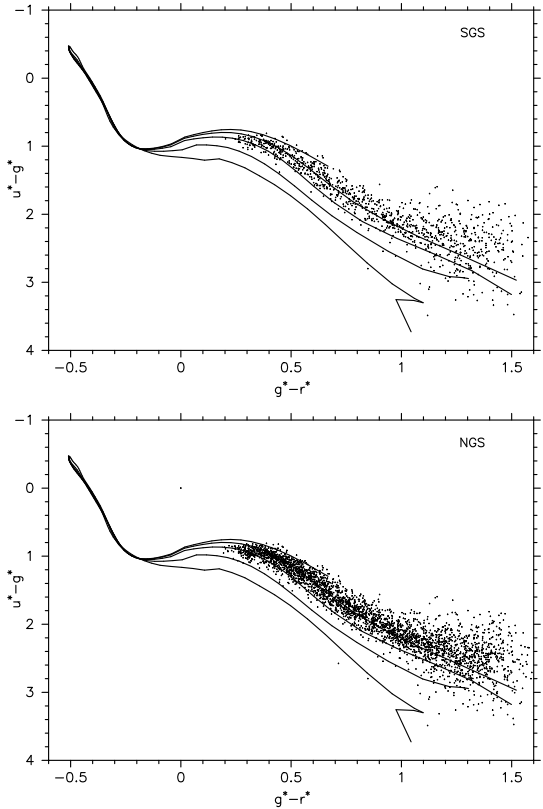}
\caption{Colour-colour plots for subdwarfs in the SGS (top) and NGS
(bottom), with model atmosphere predictions for stars of [m/H] =
$+1.0, 0.0, -1.0, -2.0$ and $-5.0$ (from bottom to top)
overlaid. There is no evidence for any systematic metallicity
difference between the samples.}
\label{fig:metal_twocol}
\end{figure}

The highly accurate five band photometry of the SDSS is ideally suited
for estimating metallicities from colours. The photometric precision
is sufficient to clearly separate stars of differing metallicities
when plotted in metallicity-sensitive colours such as ($\sdssug)$ and
$(\sdssgr)$ (Figure~\ref{fig:metal_twocol}). We use all five
photometric bands matched to model atmosphere grids in order to derive
a metallicity estimate for each subdwarf candidate.

The model atmospheres of \citet{Lenz et al.1998} use Kurucz ATLAS9
models \citep{Kurucz 1993} to give grids of four synthetic SDSS
colours for stars with effective temperatures ($T_{eff}$) from 3500 to
$40\,000$ K and varying metallicity and $\log g$. Adopting a value of
$\log g$ = 4.5 and given the observed colours, we use a chi-square
statistic and interpolate in the grids to estimate the best-fit
temperature and metallicity for each subdwarf. Stars with poor
metallicity-temperature fits are eliminated by rejecting all
candidates with a residual greater than twenty times the rms value of
the residuals of the grid fit. This leads to some 3\% of candidates
being rejected, although their highly unusual colours means that most
of these objects will not be subdwarfs.

\begin{figure}
\centering \includegraphics[width=65mm,trim=0 10 0 30]{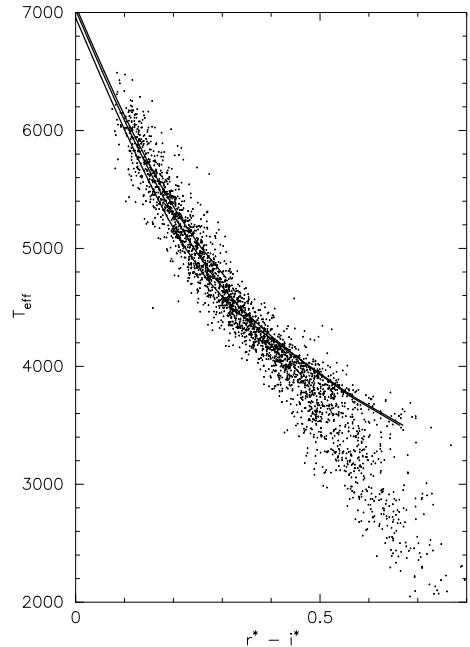}
\caption{The correlation between effective temperature and $\sdssri$,
with model predictions for [m/H] = $-1.0, -2.0, -5.0$ (in order of
increasing temperature at $\sdssri\sim0.2$) from \citet{Lenz et
al.1998}. For the majority of stars (those with $\sdssri\la0.4$) there
is close agreement with the models and good internal consistency of
the metallicity and temperature estimates. The poor fit of the models
for $\sdssri\ga0.4$ is likely to be mostly due to less accurate
temperature interpolations over this range, arising from the larger
colour spread and the fact that the [m/H] = $-5.0$ model does not
extend to these redder wavelengths. However, since the majority of the
sample lies blueward of $\sdssri\approx0.5$ and the metallicities are
not used in an absolute sense, the effect of these discrepancies will
be limited.}
\label{fig:teffcol}
\end{figure}

\begin{figure}
\centering \includegraphics[width=55mm,trim=0 10 0 30]{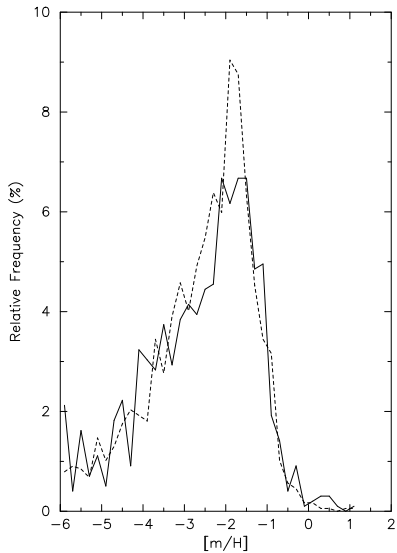}
\caption{The estimated metallicity distributions for the SGS (solid
line) and NGS (dashed line). Although the metallicities are clearly
not accurate in an absolute sense from comparison with the [Fe/H]
$\sim -1.5$ expected for spheroid stars, there is good agreement
between the samples and no systematic metallicity difference between
the two. The median metallicities of $-2.5$ for the SGS and $-2.3$ for
the NGS agree well within the expected errors.} \label{fig:metaldist}
\end{figure}

It is clear from inspection of the metallicities derived that there is
a zero point error in the estimates, with the median metallicity for
each stripe of $\widetilde{[m/H]} \sim -2.4$ considerably lower than
the [Fe/H]$\sim -1.5$ expected for spheroid stars \citep*{Laird et
al.1988, Nissen & Schuster 1991}. There are a number of likely
contributing causes of this: one is that the [m/H] = $-5.0$ model only
extends blueward of $(\sdssgr) \approx 0.7$, and so the metallicities of
subdwarfs redder than this have to be extrapolated from the [m/H]$ \ge
-2.0$ models. Even subdwarfs with $(\sdssgr) \la 0.7$ rely on
interpolation between the [m/H]$=-5.0$ and $-2.0$ models -- a very
large metallicity spread which will inevitably introduce errors.
Additionally, inaccuracies in the model atmospheres will lead to
uncertainties in the derived metallicities, particularly for redder
stars. However, the observed offset is largely irrelevant to this
application as we are interested in only the \emph{relative}
metallicities of the sample. Aside from the zero point error, the
derived metallicities and temperatures show good consistency, both
with the expected correlations on a temperature-colour diagram
(Figure~\ref{fig:teffcol}) and in colour-colour plots
(Figure~\ref{fig:metal_twocol}). Additionally, the metallicity
distributions of both the SGS and NGS are in close agreement
(Figure~\ref{fig:metaldist}), indicating no significant systematic
difference (whether intrinsic or not) between the samples.

\subsubsection{Photometric parallaxes} \label{sec:photparx}

With a metallicity estimate for each candidate subdwarf, a
metallicity-dependent colour-magnitude relation is used to provide a
more reliable indication of each star's intrinsic luminosity. A median
colour-magnitude relation $\widetilde{M_I}$ is derived by fitting a
single cubic spline to all the parallax subdwarfs in
Section~\ref{sec:cmrsel}. The absolute magnitude $M_I^*$ of each star in the
sample is then estimated by offsetting from this median magnitude by
an amount proportional to the star's deviation from the sample median
metallicity at that particular colour. The estimated absolute
magnitude is given by:
\be \label{eq:mHoffset} M_I^* = \widetilde{M_I} + d[m/H] w
\frac{dM_I}{d[m/H]}\biggr|_{(\sdssri)}, \ee
where
\be d[m/H] = [m/H]^* - \widetilde{[m/H]}_{(\sdssri)}, \ee
is the metallicity offset of the star from the sample median at that
colour, $w$ is a weighting function to allow for colour-dependent
errors, and the derivative is the variation of absolute magnitude with
metallicity derived from theoretical colour-magnitude relations.

The weight function corrects for the larger colour and metallicity
errors at redder wavelengths by assigning a weight relative to the
metallicity spread at each particular colour. Hence

\be w = \left(\frac{\sigma_{[m/H]}^{true}}{\sigma_{[m/H]}}\right)^2,
\ee where $\sigma_{[m/H]}^{true}$ is the assumed intrinsic metallicity
spread, given by the rms metallicity deviation from the median for 0.1
$\le (\sdssri) \le $ 0.25, and $\sigma_{[m/H]}$ is the rms metallicity
spread at a given colour. This weighting therefore assumes that the
metallicity range for the sample is independent of colour.

\begin{figure}
\centering
\includegraphics[width=65mm,trim=0 20 0 30]{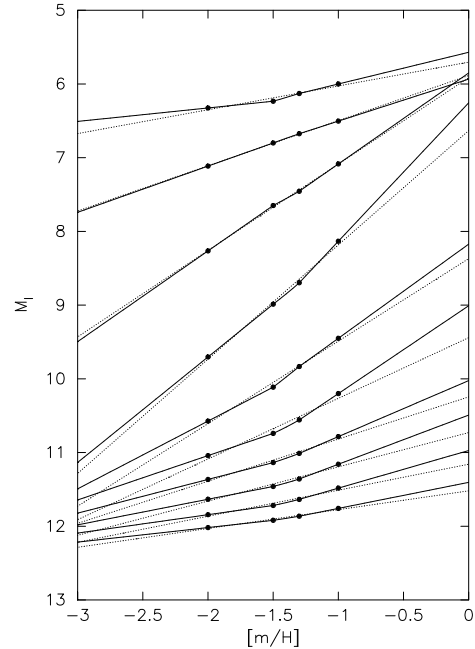}
\caption{The correlation of absolute magnitude with metallicity from
the models of \citet{BCAH97}, for $(V-I_C)$ from 1.0 (top) to 2.8
(bottom) in steps of 0.2.} \label{fig:metalmag}
\end{figure}

The derivative of absolute magnitude with respect to metallicity at
each colour is derived from the theoretical colour-magnitude relations
for stars of different metallicities provided by \citet{BCAH97}. The
metallicity-absolute magnitude relation varies considerably with
colour (Figure~\ref{fig:metalmag}), but is approximately linear. The
slope of this relation is therefore a constant and is estimated from
the models for each star's colour to provide the last term in
Equation~(\ref{eq:mHoffset}). The models of \citet{BCAH97} and the
parallax subdwarfs are given in the $(M_I, V-I_C)$ colour plane;
conversion to SDSS colours is achieved using
Equation~(\ref{eq:cmrconv}).

\begin{figure}
\centering
\includegraphics[width=65mm,trim=0 30 0 50]{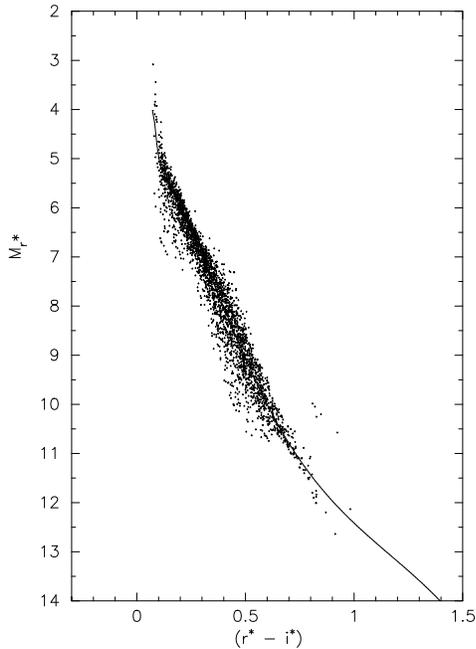}
\caption{Estimated absolute magnitudes for all subdwarfs in the SGS
and NGS, derived from a median colour-magnitude relation fitted to
calibrating subdwarfs (solid line) and offset according to the
metallicity estimate of each star.} \label{fig:cmr}
\end{figure}

It is stressed that this modification of the colour magnitude relation
based upon estimated metallicity is only an approximation. It is only
intended to adjust the predicted magnitude of each star according to
its likely metallicity in the right direction and by roughly the right
amount. In light of this the adjustment of absolute magnitude from the
median is restricted to a maximum of $\pm$1.5 mag, to prevent any
stars being assigned inappropriate values. However, it can be seen
from the sample estimated colour-magnitude diagram
(Figure~\ref{fig:cmr}) that this method provides a far more realistic
and accurate distribution of absolute magnitudes than a simple
division of the colour-magnitude relation into two metallicity ranges.

\begin{figure}
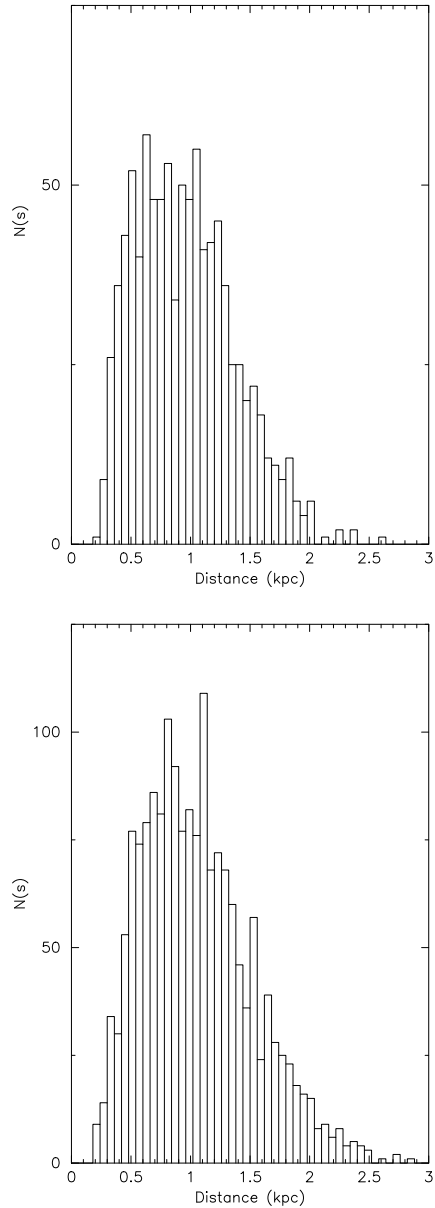

\centering
\includegraphics[width=55mm,trim=20 10 20 30]{fig17a.eps}
\includegraphics[width=55mm,trim=20 10 20 30]{fig17b.eps}
\caption{The distributions of estimated distances for the SGS (top)
and NGS (bottom).} \label{fig:distances}
\end{figure}

With an estimated absolute magnitude for each star a distance can then
be derived, assuming no significant reddening effects due to the high
latitude of the fields ($|b| \ge 37 \degr$). The distances sampled are
in the range $\sim$260 pc to $\sim$2.8 kpc, and the distributions for
the SGS and NGS are shown in Figure~\ref{fig:distances}. That
proportionally fewer stars with distances over 2 kpc are found in the
SGS compared to the NGS is an effect expected in a non-uniform
spheroid density distribution. Whereas the NGS is directed principally
towards the inner quadrants of the Galaxy, the SGS lies towards the
outer quadrants where the space density of stars will be lower with a
non-uniform distribution.

\subsection{Tests for sample systematic errors}

The luminosity functions derived from the two independent SGS and NGS
samples 
not only provide an excellent probe of the spheroid density
distribution, but also can be compared to ensure that our results are
free from significant systematic errors -- a source of concern with
earlier studies (Section~\ref{sec:recentresults}).

Prior to comparing the results from each stripe, however, we must
ensure that the two samples are self-consistent and free from
systematic effects. Inconsistency between the samples could arise
from either a systematic disparity in the selection of each
sample, or from an intrinsic difference in the constituent stars
of each.

\subsubsection{Sampling systematic errors} \label{sec:samplerr}

Systematic errors arising from the sample selection are the most
likely, with a whole range of possible causes such as a difference in
astrometric or photometric accuracy between the stripes. However, we
can test for any significant difference of this type by investigating
the completeness of each subdwarf sample. We do this in a somewhat
crude manner by comparing cumulative proper motion number counts:
assuming a uniform stellar density and that proper motion and distance
are inversely proportional ($\mu \propto d^{-1}$), a plot of log
cumulative number count ($\Sigma N$) against $\log\mu$ should have a
gradient of $-3$:
\be \label{eq:mucount} \Sigma N \propto V \propto d^3 \propto
\mu^{-3}. \ee

\begin{figure}
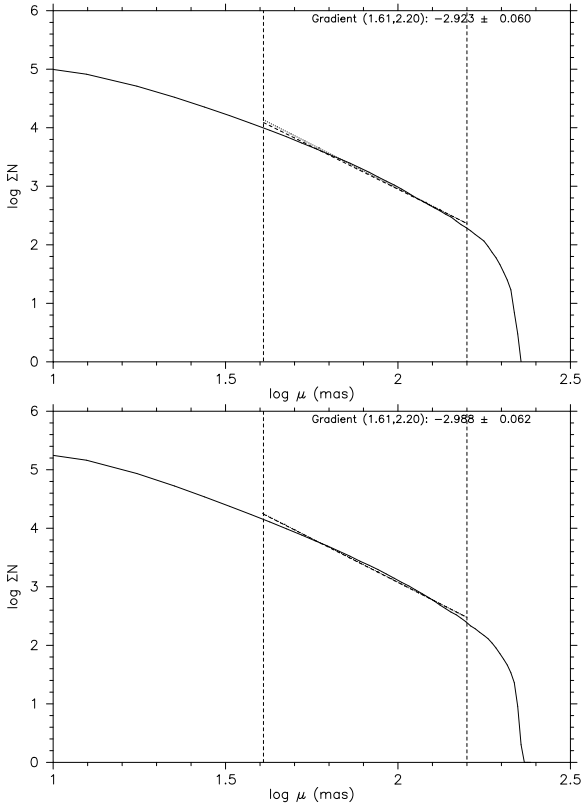

\centering
\includegraphics[height=80mm,angle=270,trim=30 0 10 0]{fig18a.eps}
\includegraphics[height=80mm,angle=270,trim=10 0 10 0]{fig18b.eps}
\caption{ Cumulative proper motion number counts for all stars in the
SGS (top) and NGS (bottom). The straight line fits between the proper
motion limits in each stripe are shown (dashed line), along with the
expected line of gradient of $-3.0$ (dotted line).} \label{fig:mucount}
\end{figure}

\begin{figure}
\centering
\includegraphics[height=85mm,angle=270,trim=30 0 10 0]{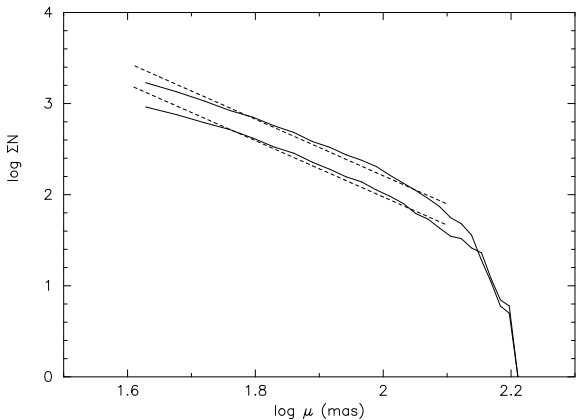}
\caption{ Cumulative proper motion number counts for candidate
subdwarfs in the SGS (lower line) and NGS (upper line). The straight
line fits between $1.6 \le\log\mu \le$ 2.1 (dashed lines) both have
gradients of $-3.1$}
\label{fig:mucount2}
\end{figure}

Figure~\ref{fig:mucount} shows the histogram of cumulative proper
motion counts for all paired stars in the NGS and SGS after the error
mapper has been applied. We fit a straight line to the points between
the global proper motion limits for each stripe: $\mu_{\rm{min}}$ =
40.5 \masyr\ and $\mu_{\rm{max}}$ = 160 \masyr.  Both give a close fit
to the expected gradient: $-2.92 \pm$ 0.06 for the SGS and $-2.99 \pm$
0.06 for the NGS. Similar plots for just the candidate subdwarfs in
each stripe are shown in Figure~\ref{fig:mucount2}, where gradients of
$-3.10$ are found for each stripe.

There is some evidence for incompleteness in the non-linearity of the
histograms, especially for the subdwarf samples where proper motion
errors lead to incompleteness towards the proper motion limits. This
is particularly relevant to the higher proper motions where the errors
will be more dominant, and this explains the rapid tailing off of the
distributions for $\log\mu \ga$ 2.1 caused by the imposition of the
upper proper motion limit. However, the number counts for the
population from which the subdwarfs are selected (Figure
~\ref{fig:mucount}) demonstrate that it is complete for 40.5 \masyr
$\le \mu \le$ 160 \masyr, and the good match of all of the histograms
to the expected gradient of $-3$ indicate that there are no significant
signs of incompleteness affecting the samples. It must be also
stressed that these are only approximate tests for completeness, and
will have some unreliability introduced by the assumption of uniform
stellar density, which is particularly invalid for stars at large
distances which tend to have lower proper motions.

\subsubsection{Radial metallicity gradient}

An intrinsic difference between the samples that could have a
significant effect on the results is a difference in median
metallicity between the SGS and NGS. Although the radial metallicity
gradient of the inner spheroid is thought to be small
($\Delta[Fe/H]/\Delta r \approx -0.012$ dex kpc$^{-1}$, \citealt{Bekki
& Chiba 2001}), with both samples in opposite radial Galactic
directions this could nevertheless introduce a source of error. The
assumption of a single colour-magnitude relation for both of the
samples would mean that stars in the sample with the higher
metallicity would have their absolute magnitudes overestimated, and
hence their distances and contribution to the luminosity function
would be underestimated. The derived luminosity function of each
sample would therefore be expected to differ. This effect would be
less pronounced with the photometric parallax relation defined in
Section~\ref{sec:photparx} that partially accounts for metallicity
variations, but is an issue that should still be addressed.

However, our data display little evidence of metallicity gradient.
The two-colour diagrams for the subdwarf sample in each stripe in
Figure~\ref{fig:metal_twocol} show no systematic difference, and the
estimated metallicity distribution and median metallicity of each
(Figure~\ref{fig:metaldist}) are very similar. It is therefore
highly unlikely that there is any strong systematic metallicity
difference between the samples.

These checks of sample consistency indicate that there are no large
systematic differences between the samples arising from sampling
errors or intrinsic variations. A further and final test of the
sample completenesses can be ascertained following the luminosity
function calculations from the value of $\langle V/V_{\rm max}
\rangle$ (Section~\ref{sec:results}).

\section{Methods: The luminosity function} \label{sec:methods_lf}

\subsection{The generalised $\bmath{V_{max}}$ method} \label{sec:vmax}

With a final sample size of 918 candidate subdwarfs in the SGS and
1696 in the NGS, the luminosity function can be accurately derived. We
achieve this using a modification of the \Vmax\ method of
\citet{Schmidt 1968}: each star contributes 1/\Vmax\ to the luminosity
function, where \Vmax\ is the maximum volume that the star could have
been detected in, given the proper motion and magnitude limits of the
survey. This technique therefore implicitly corrects for any bias
arising from the proper motion selection.

Schmidt's original 1/\Vmax\ method assumes that the sample is selected
from a uniformly-distributed population. This certainly is not the
case for our subdwarf sample; indeed, we intend to use it to determine
the variation in spheroid density. We therefore adopt the `generalised
\Vmax' method \citep*{Stobie et al.1989, Tinney et al.1993}, which
extends Schmidt's method to non-uniformly distributed samples. Given
that each star in the survey can be detected to minimum and maximum
distances ${d_{min}, d_{max}}$ in a field of solid angle $\Omega$, the
luminosity function is defined as:

\be \label{eq:samplelf} \rm{\Phi_{sample} = \sum
\frac{1}{V_{max}^{'}}}, \ee

\be \label{eq:vmax} {V_{max}^{'}} = \Omega \int^{
d_{max}}_{ d_{min}} s^2\frac{n(s)}{n_{\odot}}\ ds, \ee
where $s$ is the heliocentric distance and $n(s)$ is the adopted
spheroid density law (Section~\ref{sec:denslaw}).

\subsection{Proper motion and magnitude limits} \label{limits}

The distance limits depend on the minimum and maximum proper motion
and magnitudes in each field. The upper and lower proper motion limits
employed are $\mu_{min} = 40.5$ \masyr\ and $\mu_{max} = 160$ \masyr\
(Section~\ref{sec:pmselection}).

Ascertaining the magnitude limits for each field is less
straightforward as the data come from two quite different sources, and
there is significant variation in the depth of the plate
material. However, the fact that the datasets are paired and that the
SDSS data probe much fainter ($r^* \sim$ 23) than the POSS-I plates
means that only the $\sdssr$ magnitudes need be considered.

\begin{figure}
\includegraphics[width=55mm, angle=270, trim=10 0 10 0]{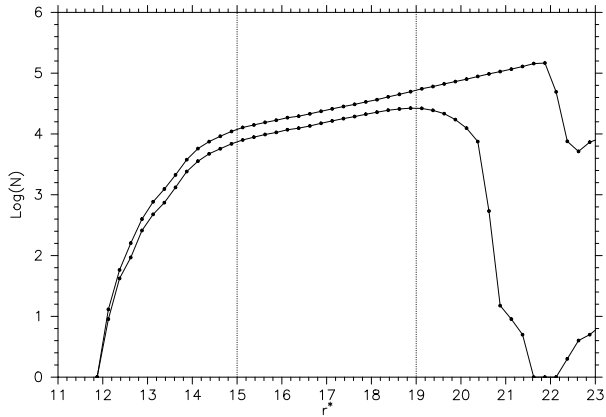}
\caption{The log number - magnitude histogram for all unpaired (top)
and paired (bottom) stars in the NGS, with the approximate upper and
lower $\sdssr$ magnitude limits for the NGS fields shown. The actual
magnitude limits adopted for each field are ascertained from where the
$\log(N(\sdssr))$ curve in the field's histogram departs from a steady
increase.}
\label{fig:maglim}
\end{figure}

The POSS-I plates reach incomplete levels at R $\sim$ 20, at which
magnitude the SDSS data are certainly complete. Stars need to appear
in both datasets to pass the pairing criteria, so the faint magnitude
limit can be defined solely in terms of the much more accurate SDSS
$r^*$ magnitude. A (log number, $\sdssr$ magnitude) histogram is
plotted for all of the paired stars in each field, and an upper limit
is conservatively defined from the magnitude at which
$\log\left(N\left(\sdssr\right)\right)$ departs from a steady increase
(Figure~\ref{fig:maglim}; Table~\ref{tab:ngslimits}.)  The brighter
limit is less crucial due to the smaller likelihood of subdwarfs
having such magnitudes, but a limit of $\sdssr\sim15$ is applied to
encompass the range that is likely to be complete as indicated by the
histogram.

\begin{table}
\centering
\caption{Magnitude limits and spheroid and thick disc discovery fractions
for SGS (top) and NGS (bottom) fields. The minimum magnitude limit for
all fields is $\sdssr$ = 15.0 and the proper motion limits are
$\mu_{min} = 40.5$ \masyr\ and $\mu_{max} = 160$ \masyr.}
\label{tab:ngslimits}
\begin{tabular}{cccc} \hline
Field & $\sdssr_{max}$ & $\chi_h$(\%) & $\chi_{td}(\%)$ \\ \hline
0932 & 18.7  & 61.15 & 0.01 \\
0363 & 19.5  & 62.56 & 0.01 \\
1453 & 19.0  & 63.57 & 0.01 \\
1283 & 19.2  & 64.16 & 0.02 \\
0852 & 19.1  & 64.39 & 0.02 \\
0362 & 19.2  & 64.27 & 0.02 \\
1259 & 19.1  & 63.85 & 0.03 \\
1196 & 19.1  & 63.13 & 0.03 \\
0591 & 19.3  & 62.13 & 0.04 \\
0319 & 19.2  & 60.85 & 0.04 \\
0431 & 19.0  & 59.31 & 0.04 \\
0834 & 18.9  & 57.50 & 0.04 \\ \hline
0151 & 19.0 & 61.15 &0.01 \\
1402 & 19.2 & 62.56 &0.01 \\
1613 & 18.8 & 63.57 &0.01 \\
1440 & 19.1 & 64.16 &0.02 \\
1424 & 19.1 & 64.39 &0.02 \\
0465 & 19.0 & 64.27 &0.02 \\
1595 & 18.7 & 63.87 &0.03 \\
1578 & 18.9 & 63.13 &0.03 \\
1405 & 19.0 & 62.12 &0.04 \\
1401 & 19.2 & 60.85 &0.04 \\
0471 & 19.2 & 59.31 &0.04 \\
1400 & 19.0 & 57.50 &0.04 \\
1397 & 18.7 & 55.42 &0.03 \\
0467 & 19.3 & 53.26 &0.03 \\
0470 & 19.1 & 50.94 &0.03 \\
1318 & 18.8 & 48.62 &0.03 \\ \hline
\end{tabular}
\end{table} \vspace*{3mm}

With proper motion limits $\mu_{\rm{min}}$, $\mu_{\rm{max}}$ and
magnitude limits $r_{\rm{min}}^*$, $r_{\rm{max}}^*$ for each field,
the limiting distances for a subdwarf of distance $d$ and magnitude
$\sdssr$ are therefore defined as \be \label{eq:dmin} \
d_{\rm{min}}=d\ \rm{max}\left[\frac{\mu}{\mu_{max}},
10^{0.2\left(r_{min}^*-r^*\right)}\right], \ee

\be \label{eq:dmax}
 d_{\rm{max}}=d\ \rm{min}\left[\frac{\mu}{\mu_{min}}, 10^{0.2\left(r_{max}^*-r^*\right)}\right].
\ee

\subsection{The spheroid density law} \label{sec:denslaw}

Recent studies of the structure of the Galactic spheroid have
indicated that its density profile is flattened and follows a
power law: an axial ratio of (c/a)$\approx$ 0.6 and $\rho(r) \sim
r^{-3}$ \citep{GFB98, Sluis & Arnold 1998, Yanny et al.2000, Ivezic
et al.2000, Chen et al.2001, Siegel et al.2002, Gould 2003}. We therefore
adopt a spheroid density law of the form
\be \label{eq:denslaw} \frac{n}{n_{\odot}} = \left(\frac{r_c^2 +
R_{0}^2}{r_c^2 + R^2 + z^2/q^2}\right)^\alpha, \ee
where $n_{\odot}$ is the local spheroid density, $\alpha$ is the
power law index and $q = (c/a)$ is the axial ratio, and we assume
$R_{0}$ = 8.0 kpc and $r_c$ = 1.0 kpc. Converting the
Galactocentric cylindrical coordinates $R$ and $z$ in terms of
heliocentric coordinates $(s,l,b)$, where $s$ is the distance from
the Sun and $l$ and $b$ are Galactic latitude and longitude:
\be \label{eq:denslaw_helio} \frac{n}{n_{\odot}} = \left(\frac{r_c^2 +
R_{0}^2}{r_c^2 + R_{0}^2 + s^2 - 2sR_{0}\cos l\cos b +
s^2 \sin^2 b \frac{1-q^2}{q^2}}\right)^\alpha. \ee

The two density law parameters $\alpha$ and $q$ are allowed to vary,
and luminosity functions are derived for a range of values to find the
$(\alpha, q)$ law which best matches the observations (see
Section~\ref{sec:bestlaw}.) A significant benefit of this approach is that
the density law variables $\alpha$ and $q$ can be fitted without
having to make an assumption about the local spheroid density
$n_\odot$.

\subsection{Discovery fraction} \label{sec:discfrac}

Given distance limits and an assumed $(\alpha, q)$ density
distribution, the sample luminosity function can then be
calculated. However, the tangential velocity cut-offs cause the sample
to exclude a given fraction of spheroid stars, so this scaling needs to be
taken into account in order to derive the true spheroid luminosity
function from the sample.

The fraction $\chi_h$ of spheroid stars expected to have 200 km
s$^{-1} \le V_T \le$ 500 km s$^{-1}$ is calculated for each field
using tangential velocity calculations as described in
Section~\ref{sec:vtcutoff}. Using our adopted velocity ellipsoids,
$\chi_h$ varies between 0.58 and 0.64 in the SGS, and 0.49 and 0.64 in
the NGS (Table~\ref{tab:ngslimits}), and scales the spheroid
luminosity function as \be \label{eq:lfscale} \rm \Phi_h^{true} =
\frac{1}{\chi_h} \Phi_h^{sample}.  \ee

At this stage any possible contamination by thick disc stars can be
considered. Assuming they are also included in the sample with $V_T
\ge 200$\kms, then the derived luminosity function will
comprise a total for spheroid and thick disc members. The sample
\emph{spheroid} luminosity function can then be calculated from the
\emph{total} (spheroid plus thick disc) by
\be \rm
\Phi_h^{sample} = \lambda_h \Phi_{h+td}^{sample},
\ee
where $\lambda_h$ is the fraction of spheroid stars in the sample. This is
given by
\be \label{eq:lambdah}
\lambda_h = \frac{1}{(\chi_{td}\left/\chi_h\right.)(n_{td}\left/n_h\right.) + 1},
\ee

where $\chi_{td}$ and $\chi_h$ are the fractions of thick disc and
spheroid stars with 200\kms\ $\le V_T \le$ 500 \kms\ and $n_{td}$ and
$n_h$ are the local number densities of thick disc and spheroid stars.
The discovery fractions $\chi_{td}$ and $\chi_h$ are known from
the calculations described in Section~\ref{sec:vtcutoff}, whilst the
relative normalisation of thick disc to spheroid stars is taken from
independent studies. Assuming a thick to thin disc density ratio
of 1:10 (\citealt{Reid et al.1995}; \citealt*{Chen et al.2001, Siegel et
al.2002}) and a combined disc to spheroid normalisation of 400:1
\citep{Chabrier & Mera 1997, Holmberg & Flynn 2000}, we adopt a
thick disc to spheroid ratio of $n_{td}$:$n_h$ = 40:1. With this
consideration of thick disc contamination, the true spheroid
luminosity function is therefore derived from
\be \label{eq:lfscale2} \rm
\Phi_h^{true} = \frac{\lambda_h}{\chi_h} \Phi_{h+td}^{sample}.
\ee

However, with a tangential velocity cut-off of $V_T \ge 200$ \kms,
this scaling for thick disc contamination has very little effect on
the luminosity function.  This normalisation gives a scaling factor of
$0.97 \le \lambda_h \le 1.00$, so at worst the thick disc
contamination has just a three percent effect on the normalisation of
the luminosity function.

\subsection{Luminosity function errors} \label{sec:lferrors} In
estimating the errors in the luminosity function we adopt the
assumption of Poissonian errors \citep{Felten 1976}:

\be \label{eq:lferror1}  \sigma_{\Phi_{\rm sample}}^2 = \sum{
\frac{1}{V_{max}^2}}. \ee Allowing for scaling:
\be \label{eq:lferror2}  \sigma_{\Phi_{\rm true}}^2 =
\left(\frac{\lambda_h}{\chi_h}\right)^2\sigma_{\Phi_{\rm sample}}^2. \ee

\subsection{Combining fields}

Whilst the luminosity functions and densities for each field are
derived separately to investigate the spheroid density profile, it is
desirable to calculate a combined luminosity function for the fields
in each stripe. Unfortunately the necessity of scaling to account for
the spheroid discovery fraction and thick disc contamination on a
field by field basis means that a total 1/\Vmax\ luminosity function
cannot be calculated for the whole sample. However, a combined
luminosity function for all of the fields in each stripe can be
derived by combining the luminosity functions for each field with a
simple weighted mean. The mean density and error for each magnitude
bin are then given by

\be \label{eq:lfwtmean} \rm
\overline{{\log \Phi}} = \frac{\sum{\log \Phi\left/\sigma_{\log
\Phi}^2\right.}}{\sum{ 1\left/\sigma_{\log \Phi}^2\right.}},\ee

\be \label{eq:lfwterror} \rm \sigma_{\overline{{\log \Phi}}}^2 =
\frac{1}{\sum{ 1\left/\sigma_{\log \Phi}^2\right.}}, \ee where the
summations are over the fields in each stripe.

\subsection{Transforming $\Phi (\Mr)$ to $\Phi (\rm
M_V)$} 
\begin{figure}
\centering
\includegraphics[width=60mm,trim=20 0 10 0]{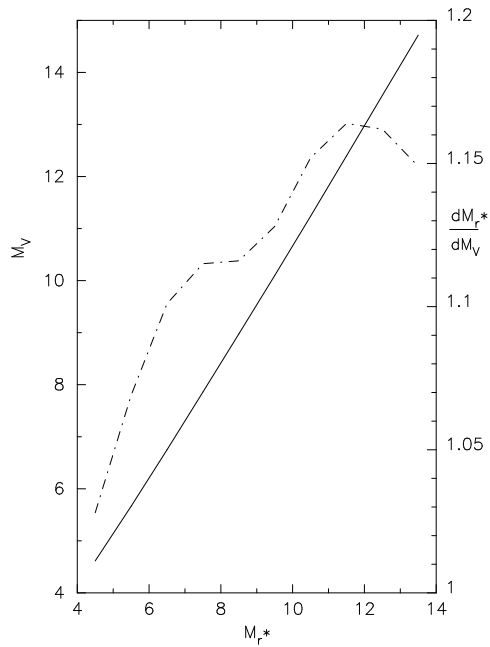}
\caption{The relations between $\Mr$ and $M_V$ (solid line) and
$\frac{dM_{\sdssr}}{dM_V}$ (dot-dashed line) derived from the
two-colour relations. These are used to convert the luminosity
function derived in $\Mr$ to $M_V$ to facilitate comparison with
published results. } \label{fig:mvmr}
\end{figure}

To facilitate comparison with published luminosity
functions, it is necessary to convert the derived luminosity
function from SDSS $\sdssr$ absolute magnitude to Johnson $V$. This
is achieved using the transformation
\be \label{eq:convlf} \Phi\left(M_V\right) =
\Phi\left(M_{r^*}\right) \frac{dM_{r^*}}{dM_V}. \ee The derivative
is evaluated from
\begin{eqnarray} \label{eq:mvmr}
M_V & = & M_{r^*} - (r^* - V) \\
    & = & M_{r^*} - \left[(-0.889(r^*-i^*)-0.040\right],
\end{eqnarray}
 from Equation~(\ref{eq:twocol1}). By varying ($\sdssri$), $\Mr$ can
be plotted against $M_V$, a spline fitted and the derivative evaluated
(Figure~\ref{fig:mvmr}). Hence $dM_{r^*}/dM_V$ can be calculated for a
given $\Mr$, and repeating this for each $\Mr$ bin in the
luminosity function completes the transformation.

\subsection{Constraining the spheroid density profile}
\label{sec:bestlaw}

\begin{figure}
\centering
\includegraphics[height=110mm]{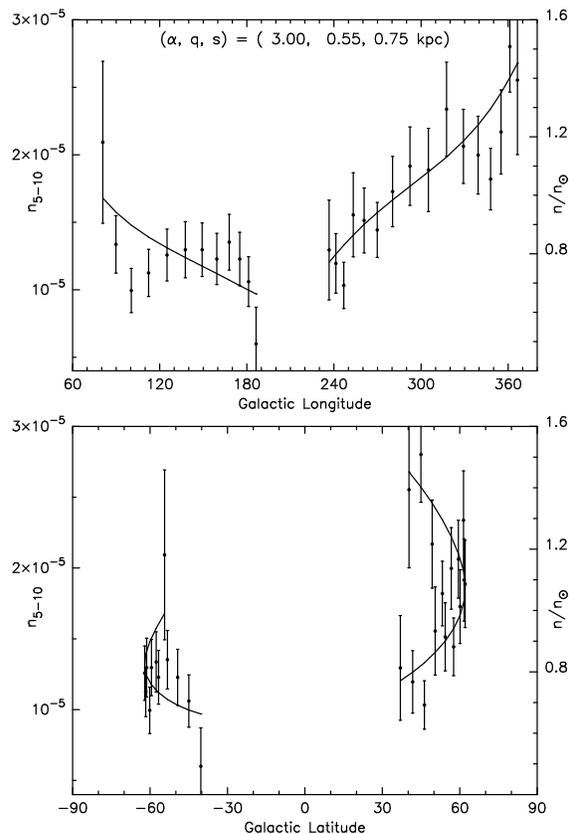}
\caption{The subdwarf number densities (points) in each field derived
from a conventional \Vmax\ method assuming uniform population
density. The lines show the expected relative number densities at a
distance of 1kpc from a spheroid density power law with
$(\alpha,q)=(-3.0, 0.55)$ . Is is clear that the uniform density
assumption is invalid for this sample and so a non-uniform
distribution must be assumed in the calculation of the luminosity
functions. Accordingly, these data can be used to constrain the form
of the density profile.}
\label{fig:lfuniform}
\end{figure}

With a wide range of lines of sight, the SGS and NGS samples are
ideally suited to constraining the density distribution of the
spheroid. Indeed, a non-uniform density law has to be assumed in order
to compare the luminosity functions of the different samples. Figure
\ref{fig:lfuniform} shows the number densities (luminosity function
integrated over 5 $\le \Mr \le$ 10) for each field plotted against
Galactic longitude when derived using a conventional \Vmax\ method
under the assumption of uniform space density. It is clear from
comparison with the models that at the distances sampled the varying
density of the spheroid has a strong effect and so has to be taken
into account when deriving the luminosity function.

In order to constrain the density distribution a power law of the
form given in Equation~(\ref{eq:denslaw_helio}) is assumed, and subdwarf
number densities are calculated for each field in the SGS and NGS
for a range of power law indices and axial ratios. The power law
index $\alpha$ is allowed to vary from $-2.0$ to $-4.4$ in steps of
0.05, and the axial ratio from 0.2 to 1.0 in intervals of 0.05.

The generalised \Vmax\ method (Section~\ref{sec:vmax}) should ensure
that the derived number densities are constant and independent of line
of sight, so how well a $(\alpha, q)$ power law model fits the data is
ascertained by measuring how close to a uniform value the field number
densities are under the model. This goodness of fit is defined by a
modified chi-square statistic

\be  \label{eq:chisq} \chi^2 = \sum_{i=1}^{nf}\sum_{j=i}^{nf}
\frac{\left(n_i - n_j\right)^2}{\sigma_i^2 + \sigma_j^2}, \ee
\noindent where $n_i$ and $\sigma_i$ are the number density and its
standard deviation for each of $nf$ fields. The best-fit $(\alpha, q)$
density model is therefore the model with the minimum value of
$\chi^2$, and the likelihood of each model can be evaluated by
comparing its statistic value with the minimum.

\section{Results and discussion} \label{sec:results}

\subsection{Spheroid density profile} \label{sec:densresult}

\begin{figure}
\centering
\includegraphics[width=75mm,trim=0 10 0 0]{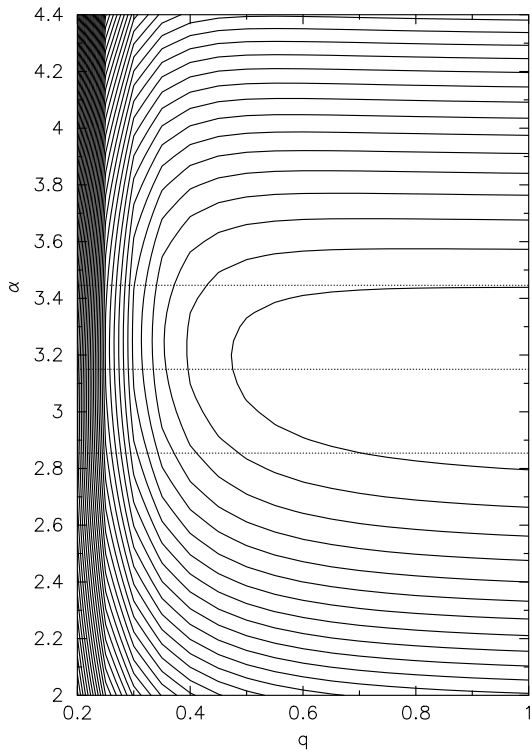}
\caption{Contours of equal $\chi^2$ constraining the spheroid density
models $(\alpha, q$). Although $q$ cannot be adequately constrained
with these data, we find $\alpha = -3.15 \pm$ 0.30. The dashed lines
show the one standard error interval from the best-fit $\alpha$. The
standard error in $\alpha$ is calculated from bootstrap resampling.}
\label{fig:denscontour}
\end{figure}

Figure~\ref{fig:denscontour} shows the results of fitting spheroid
density models $(\alpha, q)$ to the combined SGS and NGS data, with
contours of equal $\chi^2$ plotted. Whilst the axial ratio $q$ cannot
be constrained with these data, the power law index $\alpha$
can. Although there is a slight degeneracy with $q$, the best-fit
power law index is $\alpha = -3.15 \pm$ 0.30 for the range 0.55 $\le q
\le$ 0.85, which is the value of the axial ratio derived from recent
studies \citep*{Sluis & Arnold 1998, GFB98, Robin et al.2000, Chen et
al.2001, Siegel et al.2002}. This power law index is largely in
agreement with the recent results of \citet{GFB98} ($\alpha = -3.13 \pm
0.23$); \citet{Sluis & Arnold 1998} ($\alpha = -3.2 \pm 0.3$);
\citet*{Yanny et al.2000} ($\alpha = -3.2 \pm 0.3$); \citet*{Ivezic et
al.2000}($\alpha = -3$); \citet*{Chen et al.2001} ($\alpha = -2.55 \pm
0.3$); \citet*{Siegel et al.2002} ($\alpha = -2.75$) and \citet{Gould
2003} ($\alpha = -3.1 \pm 1.0$).  This $r^{-3}$ distribution differs
significantly from the $r^{-2}$ power law of the dark matter halo,
providing further evidence for the theory that faint halo stars
constitute at most a negligible fraction of Galactic dark matter
\citep{Bahcall et al.1994, Flynn et al.1996, Chabrier & Mera 1997,
Fields et al.1998}.

These results are only able to constrain $q \ga$ 0.3. Comparison of
the limits of $q \approx 0.55 \pm 0.06$ set by \citet{Chen et al.2001}
with the SDSS EDR data and inspection of
Equation~(\ref{eq:denslaw_helio}) shows that this is due to the much
smaller distances probed by this study. Small values of $s$ in
Equation~(\ref{eq:denslaw_helio}) cause the $q$ term to be negligible
compared to the $r_c^2 + R_0^2$ terms, and so variations of $l$ and
$b$ are insensitive to $q$. The low distances are a result of the
relatively bright lower magnitude limit of $\sdssr\sim19$ set by the
POSS-I R plates; using a wider range of Galactic coordinates from
future SDSS releases will help to overcome this degeneracy.

A similar problem afflicts independent analysis of the density
distributions of the SGS and NGS. The smaller ranges of Galactic
longitude and distances (Figure~\ref{fig:distances}) sampled in the
SGS and the decreased sensitivity to the density profile parameters in
this direction means that neither $\alpha$ or $q$ can be adequately
constrained for the SGS taken in isolation. The NGS alone does allow
limits to be placed on $\alpha$ and $q$, and these agree well with the
values derived for the joint SGS/NGS sample, albeit with larger
errors.

The standard error of the $\alpha$ derived for the SGS and NGS is
obtained from bootstrap resampling. The best-fit $\alpha$ is
calculated for each of 1000 bootstrap samples, all of the same size as
the original sample and chosen from it by randomly selecting stars
with replacement. The standard deviation of $\alpha$ for these
bootstrap samples is an estimate of the standard error in the power
law index.

\subsection{The subdwarf luminosity function} \label{sec:lfresult}

\begin{figure*}
\centering \includegraphics[width=130mm,angle=270,trim=0 70 0 30]{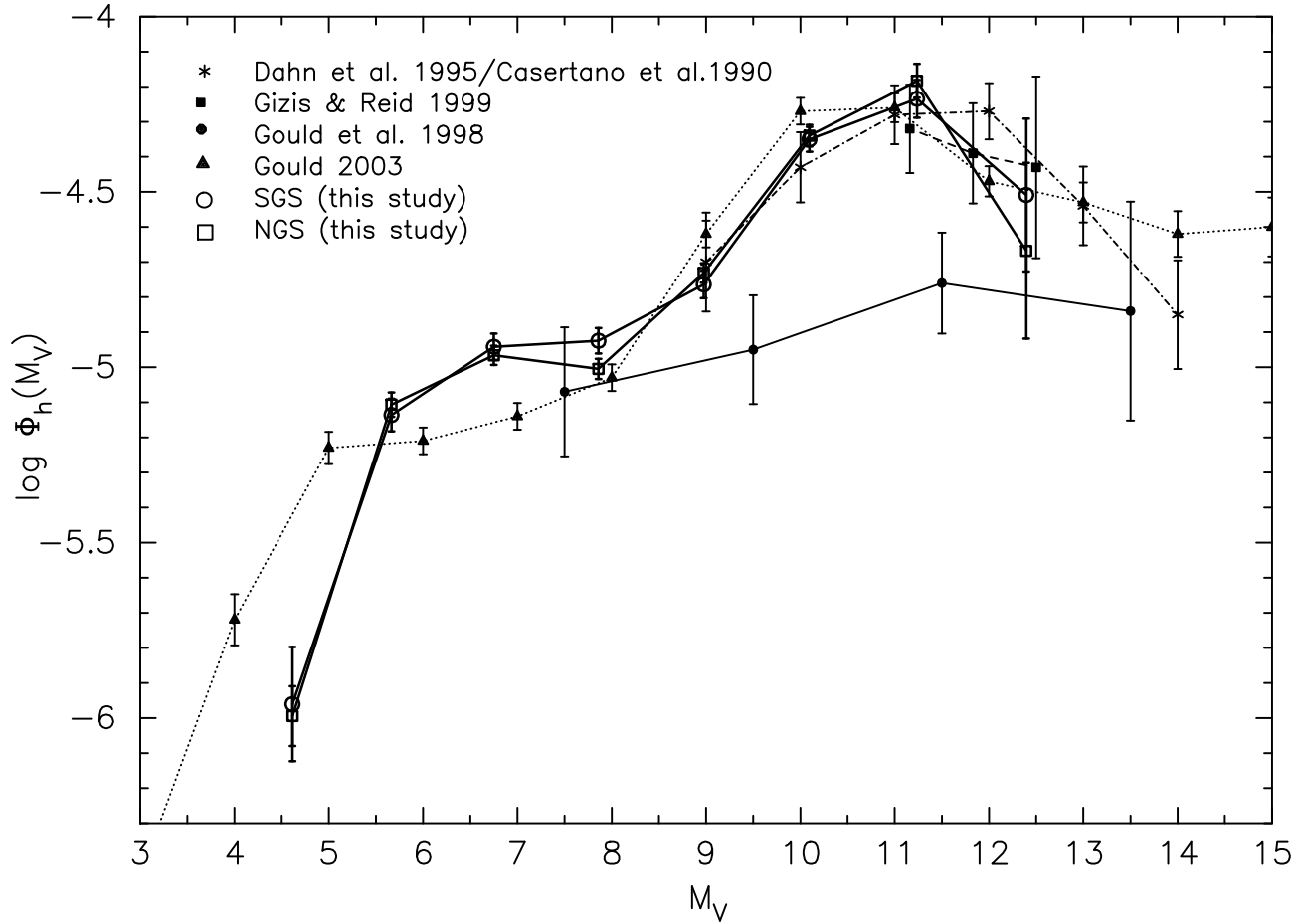} 
\caption{The combined luminosity function for the SGS (solid line,
open circles) and NGS (solid line, open squares) assuming the best-fit
spheroid density power law $\alpha = -3.15$. The SGS and NGS results
are in excellent agreement with each other and agree well with the
results of \citet{D95}/\citet{Casertano et al.1990} (dot-dash,
asterisks), \citet{GR99} (dashed, filled squares) and \citet{Gould
2003} (dotted, filled triangles), but not with the outer spheroid
sample of \citet{GFB98} (solid line, filled circles). This lends
weight to the suggestion that a single power law cannot describe the
density distributions of both the inner and outer spheroid. [Note that
the \citealt{D95} luminosity function has been scaled by 0.75 to
reflect use of the \citealt{Casertano et al.1990} kinematics, as in
\citealt{GFB98}.] The small error bars of this study's luminosity
function reflect the much greater size of the subdwarf sample
compared to the other kinematic studies; \citet{Gould 2003} has a
larger sample size, although his is more likely to suffer from thick
disc contamination.}
\label{fig:lfresult}
\end{figure*}

The combined luminosity functions of the SGS and NGS fields for the
best-fit power law $\alpha = -3.15$ are shown in Figure
\ref{fig:lfresult} (the luminosity functions for our samples are
insensitive to the value of $q$.) The results for each stripe are in
excellent accordance with each other, with all but one magnitude bin
agreeing within the 1$\sigma$ error bars. This suggests that there are
no systematic spatial effects in the analysis of the two samples.

There is also good agreement with the kinematic studies of
\citet{GR99}, \citet{D95} (scaled by 0.75 to account for
\citealt*{Casertano et al.1990} kinematics as in \citealt{GFB98}) and
\citet{Gould 2003}. This indicates that the correction of our sample
to the solar neighbourhood data by an $r^{-3.15}$ power law is a good
approximation, and that this distribution therefore well describes the
inner spheroid out to 2.5kpc. The disagreement with the result of
\citet{GFB98} is pronounced, however, and this may be due to
systematic errors affecting their sample: for example, \citet{GR99}
postulated that the choice of local calibrating subdwarfs in
\citet{GFB98} could cause their luminosity function to be
underestimated. Alternatively, there could be a real effect behind the
discrepancy. The inability of the \citet{GFB98} data to match the
local luminosity functions when fitted with a $(\alpha, q) = (-3.13,
0.8)$ density power law indicates that this model does not accurately
describe the spheroid. As we find that an $r^{-3.15}$ power law
accurately fits our data out to 2.5kpc, this provides more evidence
that a difference in the axial ratios of the inner and outer spheroid
\citep{Sommer-Larsen & Zhen 1990} may be responsible for the
discrepancy of the \citet{GFB98} result. However, there is growing
evidence that the usual four-component model for the Galaxy (with
discrete thin and thick discs, halo and bulge) is inadequate, and that
a model with more continuous distributions in stellar age, kinematics
and metallicity is more appropriate \citep{CB2000, Siegel et al.2002,
Yanny et al.2003}. It is possible that the deficiencies of this
over-simplistic picture of Galactic structure is behind the
discrepancies seen here.

A further feature evident from Figure~\ref{fig:lfresult} is the
precision of our results. That our error bars are so small reflects
the size of our sample, with only the \citet{Gould 2003} luminosity
function exhibiting comparable errors. However, our subdwarf selection
is more rigorous that of \citet{Gould 2003}, who identified $\sim$4500
subdwarfs simply by taking cuts by eye in the reduced proper motion
plane and whose sample is perhaps more susceptible to thick disc
contamination.

\begin{figure}
\centering
\includegraphics[width=75mm,trim=0 20 0 0]{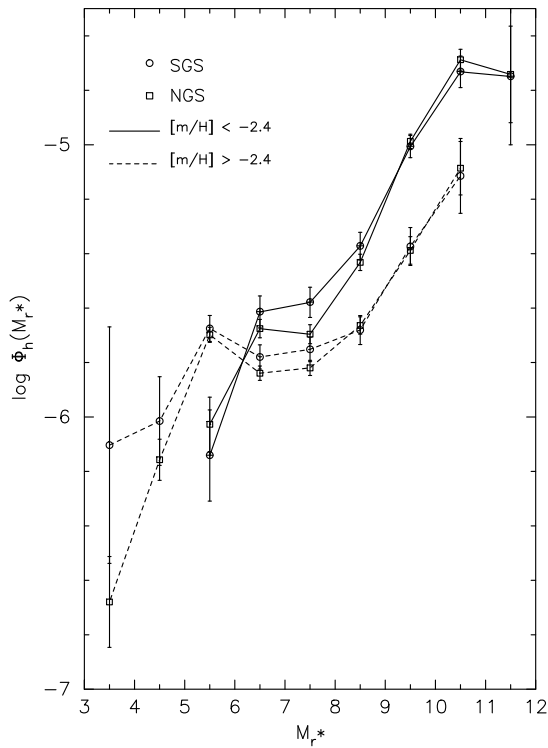}
\caption{The luminosity functions for stars with metallicities either
greater (dashed line) or less than (solid line) the median metallicity
of $[m/H] \approx -2.4$ in the SGS (circles) and NGS (squares). As in
\citet{GR99}, we find that more metal-poor subdwarfs have a higher
space density.}
\label{fig:lfmHsplit}
\end{figure}

Figure~\ref{fig:lfmHsplit} plots the luminosity functions for stars of
different metallicities, with the samples divided into subdwarfs with
metallicities either greater or less than the median of
$\widetilde{[m/H]} \approx -2.4$. As in the study of \citet{GR99}, we
find that the very metal-poor subdwarfs extend to fainter absolute
magnitudes and tend to have a higher space density than the more
metal-rich ones.

\subsection{The $\bmath{\langle V/V_{\rm max}\rangle}$ test} \label{sec:vvmax}

The overall completeness of the subdwarf samples can be estimated by
using the $\langle V/V_{\rm{max}}\rangle$ test. For each star the
ratio of the volume $V$ (corresponding to its distance $d$) to $V_{max}$
is calculated (Equation~\ref{eq:vvmax}), and the mean of this quantity
should be 0.5 for a complete survey evenly sampling the survey volume.
The error in this mean is $1/\left(12N\right)^{\frac{1}{2}}$, where N
is the number of stars in the sample.

\be \label{eq:vvmax} 
\left\langle\frac{V}{V_{max}}\right\rangle = \left\langle\left(\frac{d}{d_{max}}\right)^3\right\rangle.
\ee

\begin{figure}
\centering
\includegraphics[height=85mm,angle=270,trim=0 60 0 0]{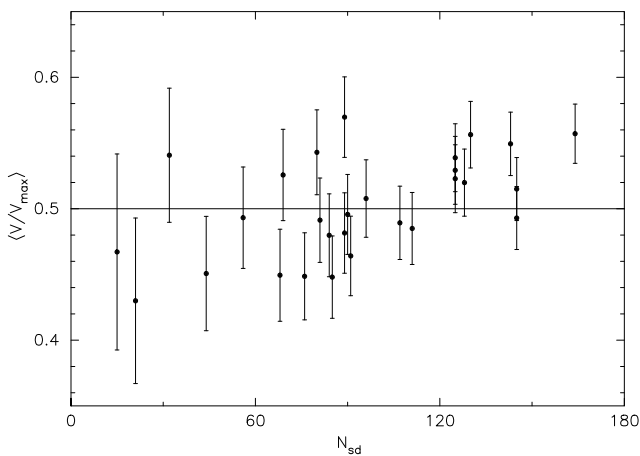}
\caption{ $\langle V/V_{\rm{max}}\rangle$ for fields in the SGS and
NGS . There is no evidence for incompleteness in the samples.}
\label{fig:vvmax}
\end{figure}

The values of $\langle V/V_{\rm{max}}\rangle$ for each field in the
SGS and NGS are plotted against Galactic longitude in Figure
\ref{fig:vvmax}. The SGS has a combined value of $\langle
V/V_{\rm{max}}\rangle = 0.495 \pm$ 0.010 and the NGS has $0.506 \pm
0.007$. Although this is not a rigorous test of completeness,
especially for non-uniformly distributed samples, these results
nevertheless provide evidence that no significant incompleteness
affects the samples.

\section{Summary and future work} \label{sec:summary}

We have demonstrated the effectiveness of the method of reduced proper
motion at selecting spheroid stars, and have used it to obtain one of the
largest samples of known spheroid subdwarfs. From this we derive the
subdwarf luminosity functions to unprecedented accuracy in two
diametrically opposite lines of sight in the Galaxy. 

The large samples along different lines of sight in this study have
enabled us to constrain the form of the spheroid density distribution,
out to heliocentric distances of 2.5 kpc, to a power law with an index
of $\alpha = -3.15 \pm 0.3$. This is in accordance with other recent
results \citep{GFB98, Sluis & Arnold 1998, Yanny et al.2000, Ivezic et
al.2000, Chen et al.2001, Siegel et al.2002}. We are unfortunately
unable to adequately constrain the spheroid flattening parameter $q$
with this study (we can only rule out a very flat spheroid), due to
insufficient survey depth.

Our luminosity functions agree well with other recent local
derivations, so that with this result the solar neighbourhood subdwarf
luminosity function is now well defined to $M_V \approx$ 12.5. Our data
corrected by an $r^{-3.15}$ spheroid density distribution closely
match the local luminosity functions, suggesting that this power law
well describes the inner spheroid density profile. On the other hand,
our result further confirms the discrepancy between the local
luminosity functions and the outer spheroid luminosity function of
\citet{GFB98} corrected by an $r^{-3.1}$ power law. This provides
additional evidence that either the \citet{GFB98} result is affected by
systematic errors, or that the inner and outer spheroid follow quite
different density distributions.

A progression of this study to larger volumes will enable limits to be
placed on $q$ and will provide stronger constraints on
$\alpha$. Adding the SIRTF First Look field from the EDR will
immediately increase the sample size by $\sim$18\%, and will provide
additional lines of sight, important for determining
$\alpha$. Progressive SDSS data releases will significantly increase
the scientific return of this work. With the first formal SDSS data
release (DR1) some 20\% of the photometry is now available, increasing
to $\sim$50\% in January 2004. This represents a huge increase on the
5\% from the EDR, and will mean that tens of thousands of subdwarfs
will be found with the methods described in this paper. These will
yield even more accurate estimates of the subdwarf luminosity function
and will provide much tighter constraints on the spheroid density law
through the increase in sample size and lines of sight.

A further development of this study that is already under way is to
obtain spectra of the candidate subdwarfs.  Confirmation of their
spectral type is important for determining the accuracy of the RPM
selection methods, and for ascertaining the true level of thick disc
contamination in the samples. In addition, radial velocities will
enable the full 3D angular momentum properties of these spheroid stars
to be investigated over large scales, with accuracy sufficient for the
detection of spheroid kinematic substructure \citep*[see eg.][]{Helmi
et al.1999, Helmi & de Zeeuw 2000}. Spectra of several hundred
candidates have already been obtained using the 6dF multi-fibre
spectrograph on the UKST at AAO.

A subsequent step to be made in the near future is to convert the
luminosity functions into an accurate subdwarf mass function by use of
a mass-magnitude relation. As described in Section~\ref{sec:intro}, this has
a number of important applications, and is particularly important to
the theories of star formation and evolution. Future analyses will
also take into consideration the expected binarity fraction of the
samples in the derivation of the luminosity functions.

This work clearly demonstrates the great potential of combining
the old-style photographic surveys with the newer CCD programmes.
This combination makes optimum use of both datasets, utilising the
accurate astrometry and long time baselines available from surveys
such as the SSS, whilst taking advantage of the accurate CCD
photometry from such studies as the SDSS.

\vspace*{5mm}
{\noindent \it Acknowledgements}

We are very grateful to the referee, Raffaele Gratton, for prompt and
helpful suggestions, and to Andy Gould and Gilles Chabrier for useful
comments. A.P.D is supported by a UK PPARC studentship, and this work
makes use of the SuperCOSMOS Sky Survey
(http://www-wfau.roe.ac.uk/sss/index.html) and the Sloan Digital Sky
Survey Archive.

Funding for the creation and distribution of the SDSS Archive has been
provided by the Alfred P. Sloan Foundation, the Participating
Institutions, the National Aeronautics and Space Administration, the
National Science Foundation, the U.S. Department of Energy, the
Japanese Monbukagakusho, and the Max Planck Society. The SDSS Web site
is http://www.sdss.org/.

The SDSS is managed by the Astrophysical Research Consortium (ARC) for
the Participating Institutions. The Participating Institutions are The
University of Chicago, Fermilab, the Institute for Advanced Study, the
Japan Participation Group, The Johns Hopkins University, Los Alamos
National Laboratory, the Max-Planck-Institute for Astronomy (MPIA),
the Max-Planck-Institute for Astrophysics (MPA), New Mexico State
University, University of Pittsburgh, Princeton University, the United
States Naval Observatory, and the University of Washington.


\end{document}